\documentclass[]{aastex631}
\usepackage{amsmath}

\newcounter{appendixfigure}
\usepackage{threeparttable}
\usepackage{footnote}
\usepackage{tabularx}
\usepackage[figuresright]{rotating}
\usepackage{appendix}

\begin{document}

\title{The X-ray Emission Reveals the Coronal Activities of Semi-detached Binaries}

\correspondingauthor{Jianfeng Wu}
\email{wujianfeng@xmu.edu.cn}

\author[0000-0002-7600-1670]{Junhui Liu}
\affiliation{Department of Astronomy, Xiamen University, Xiamen, Fujian 361005, People's Republic of China}
\affiliation{Research School of Astronomy \& Astrophysics, Australian National University, Cotter Rd., Weston, ACT 2611, Australia}

\author[0000-0001-7349-4695]{Jianfeng Wu}
\affiliation{Department of Astronomy, Xiamen University, Xiamen, Fujian 361005, People's Republic of China}



\begin{abstract}

X-ray emission is an important tracer of stellar magnetic activity. 
We carried out a systematic correlation analysis for the X-ray luminosity $\log L_{\rm X}$, bolometric luminosity $\log L_{\rm bol}$, and X-ray activity level $\log(L_{\textrm{X}}$/$L_{\textrm{bol}})$ versus the binary parameters including orbital period $P$, Rossby number $R_{\rm O}$, effective temperature $T_{\rm eff}$, metallicity [Fe/H] and the surface gravity $\log g$, and the stellar mass $M$ \& radius $R$, by assembling a large sample of semi-detached (EB-type) binaries with X-ray emission (EBXs).
The fact that both $\log L_{\rm X}$ and $\log L_{\rm bol}$ change in accordance with $\log P$ indicates that X-ray emission originates from the convection zone, while $\log L_{\rm X}$ is proportional to the convection zone area. We found that EBXs with main-sequence components exhibit an upward and then a downward trend in both the $\log T_{\rm eff}$-$\log L_{\textrm{X}}$ and $M$-$\log L_{\textrm{X}}$ relations, which is different from the monotonically decreasing trend shown by EBXs containing sub-giant and giant components. The magnetic activity level is negatively correlated with $\log T_{\rm eff}$ and stellar mass. Based on the magnetic dynamo model, the variations in the size and thickness of the surface convection zones can explain the observed relations.
EBXs with main-sequence components have similar $R_{\rm O}$-$\log(L_{\textrm{X}}/L_{\textrm{bol}})$ relationship to that of the binaries in the clusters as Praesepe and Hyade. 
We compared the X-ray radiation properties of EBXs with those of the X-ray-emitting contact binaries and found that EBXs have broader value ranges for $\log L_{\rm X}$ and $\log(L_{\textrm{X}}$/$L_{\textrm{bol}})$.

\end{abstract}


\section{Introduction} \label{sec:intro}

In general, an EB-type binary, also known as a $\beta$ Lyrae-type binary, is a semi-detached close binary system with only one component filling its Roche lobe. Their typical spectral types range from late A to K \citep{2019ApJS..244...43Z}. The variable amplitudes of EB-type binaries are generally smaller than 1 magnitude, while their orbital periods span from 0.2 to several days \citep{2019ApJS..244...43Z}. The light curves exhibit fairly smooth and continuous eclipses. Furthermore, the luminosities at the two maxima are practically identical, whereas those at the minima differ considerably. In this work, our classification for the EB-type binary follows the All-Sky Automated Survey for Supernovae (ASAS-SN) survey \footnote{https://asas-sn.osu.edu/atlas/EB} \citep{2018MNRAS.477.3145J, 2021MNRAS.503..200J}. EB systems include X-ray emitters \citep{2008AcA....58..405S}, which we refer to as EBXs in this work. By combining eclipsing binaries from the ASAS with the ROSAT All Sky Survey (RASS), \citet{2008AcA....58..405S} compiled a catalog with 266 X-ray-emitting EB binaries and expanded the coronal activity study from contact (EW-type) close binaries to semi-detached close binaries. Moreover, for a given optical color, the activity level of EB-type binaries is generally higher than that of contact binaries.

For late-type main-sequence (M- to F-types) single stars, previous studies (e.g., \citealt{2016MNRAS.463.1844S, 2018MNRAS.479.2351W, 2019A&A...628A..41P, 2020ApJ...902..114W, 2020A&A...638A..20M, 2022A&A...661A..29M}) on the magnetic activity-rotation relations have revealed the relationship between X-ray emission and the stellar dynamo. The standard stellar dynamo located in the tachocline is powered by convection and rotation, connecting the solidly rotating radiative interior with the differentially rotating convective envelope \citep{1955ApJ...122..293P, 1993ApJ...408..707P}. The X-ray emission can act as a proxy for the efficiency of the stellar dynamo \citep{2022A&A...661A..29M} and serve as a manifestation of magnetic activity in the outermost atmospheric layer, namely the corona. The stellar dynamo magnetic activity produced by fast rotation and envelope convection, as well as the large-scale horizontal flow between the components, are considered to be the possible X-ray emission mechanism of EW-type close binaries (e.g., \citealt{2001A&A...370..157S, 2004A&A...415.1113G, 2006AJ....131..990C, 2022A&A...663A.115L}). 
However, there is currently limited statistical research on the relationship between the X-ray emission properties of EBXs and the stellar dynamo model. Moreover, for late-type main-sequence stars and close binaries, the ratio of X-ray luminosity $L_{\textrm{X}}$ to bolometric luminosity $L_{\textrm{bol}}$ is used to represent the magnetic activity level of an individual system \citep{1993ApJ...410..387F, 2004A&ARv..12...71G,2006AJ....131..990C,2022A&A...663A.115L}. This ratio tends to reach a maximum level at $10^{-3}$, i.e., $\log$($L_{\textrm{X}}$/$L_{\textrm{bol}}$) $\sim -3$, which is called the {\it saturation limit} or {\it saturation level} \citep{1984A&A...133..117V, 1987ApJ...321..958V, 1993ApJ...410..387F}. Studying the X-ray radiation properties of EB-type binaries can advance our understanding of magnetic dynamo models of stellar activity. 

In this study, we conduct the first systematic population study of EBXs for their coronal activities, primarily using samples selected from the ASAS-SN Variable Star catalog and the X-ray databases of XMM-Newton, RASS, and Chandra X-ray Observatory. The remainder of this study is structured as follows: Sections~\ref{sec:dataselection} and \ref{sec:data_analysis} describe the data selection and statistical analyses of EBX samples, respectively. Section~\ref{sec:Discu} discusses the relationships among the period, stellar atmospheric parameters, and magnetic activity. Section~\ref{sec:Summary} summarizes the main results.

\section{Sample Selection} \label{sec:dataselection}

\subsection{EBXs in 4XMM-DR11}\label{sec:Sample_selection_XMM}

The EB-type binaries were first selected from the databases of ASAS-SN, which periodically scans the entire visible sky with a cadence of $\sim$2$-$3 days and a sensitivity limit of $V$ $\lesssim$ 17 mag \citep{2018MNRAS.477.3145J, 2021MNRAS.503..200J}. Since 2018, the monitor of the sky has expanded to a depth of $g$ $\lesssim$ 18.5 mag with $\sim$ 1 d cadence. ASAS-SN identifies new variable star candidates by applying a random forest classifier to the light curve characteristics \citep{2019MNRAS.486.1907J}. Until November 2023, the ASAS-SN Variable Star Database (AVSD)\footnote{https://asas-sn.osu.edu/variables} lists 25932 EB-type binaries classified from $\sim$~680000 variable stars. By cross-matching variable stars with different external catalogs \citep{2023MNRAS.519.5271C}, such as $Gaia$ EDR3 \citep{2021A&A...649A...1G}, 2MASS \citep{2006AJ....131.1163S} and ALLWISE \citep{2010AJ....140.1868W}, AVSD provides information about the $Gaia$ DR3 IDs,\footnote{The AVSD had cross-matched with \textit{Gaia} EDR3. Considering that \textit{Gaia} EDR3 and DR3 have identical ID numbers and full astrometric solutions, we consistently refer to DR3 throughout this work. All stellar parameters related to \textit{Gaia} are retrieved from the DR3 catalog.} eclipsing periods, proper motion, photometry, and color/reddening for most sources. By combining the parallax information of $Gaia$ DR3, we calculated the distance of each system and eliminated those with distance $>2$~kpc or uncertainty of distance $>20$\% (parallax error/parallax $>20$\%). We selected objects within the 2~kpc distance to ensure accurate X-ray luminosity calculations, as $Gaia$ DR3 provides most reliable distances up to 2~kpc (see Section 3.1 of \citealt{2023A&A...674A..28F}). After these criteria were applied, 15039 EB-type binaries (ASAS-SN-EB) were selected as the primary catalog. 

The 4XMM-DR11 catalog contains 895415 unique X-ray sources detected during 12210 pointed \textit{XMM-Newton} EPIC observations \citep{2020A&A...641A.136W}. The high sensitivity and random nature of the observations of 4XMM-DR11 make it suitable for searching for the X-ray counterparts of EB-type binaries. \citet{2022A&A...663A.115L} verified the completeness of \textit{XMM-Newton} and \textit{RASS} samples for the study of eclipsing binaries. 

We cross-matched the ASAS-SN-EB catalog with the 4XMM-DR11 full catalog with a matching radius of $15^{\prime\prime}$. This process yielded 180 closest unique X-ray sources for the \textit{XMM-Newton} detection. To further purify the sample, we visually examined the AVSD light curves with reference to the shape of the light curves given by ASAS-SN,\footnote{https://asas-sn.osu.edu/atlas/EB} and eliminated sources that did not exhibit EB-type light curve characteristics.  A total of 150 subjects remained in the sample. The X-ray fluxes in the 0.2-12.0~keV band in the 4XMM-DR11 catalog were calculated by assuming a power-law model with a photon index of 1.42 and the hydrogen column density ($N_{\rm H}$) of 1.7$\times 10 ^{20}$cm$^{-2}$. We derived the $N_{\rm H}$ value for each object using the extinction $A_{V}$ obtained from $Starhorse$ and the relation of $N_{\rm H}$ (cm$^{-2}$) = (2.21$\pm$0.09)$\times$10$^{21}$ A$_{V}$ (mag; \citealt{2009MNRAS.400.2050G}). The average $N_{\rm H}$ value is 1.41$\times$10$^{21}$~cm$^{-2}$, which is higher than the above value adopted by 4XMM-DR11. Therefore, we re-calculated the flux values for each binary using its $N_{\rm H}$ values with PIMMS. The flux error includes the uncertainty due to the choice of the spectral fitting model.

\subsection{EBXs in RASS} \label{sec:sec:Sample_selection_RASS}

We applied the same data screening procedures as described in Section~\ref{sec:Sample_selection_XMM} to select 96 EBXs from the Second RASS source catalog (2RXS; \citealt{2016A&A...588A.103B}), except using a $20^{\prime\prime}$ matching radius. The X-ray flux in the 0.1$-$2.4 keV band was converted from the count rate using the energy conversion factor calculated from the hardness ratios provided by 2RXS \citep{1996A&A...310..801H}. In this catalog, the $N_{\rm H}$ value for each source based on \citet{1990ARA&A..28..215D} was applied.
Using $PIMMS$\footnote{https://cxc.harvard.edu/toolkit/pimms.jsp}, we transformed the unabsorbed X-ray fluxes from the 0.1$-$2.4 keV band to 0.2$-$12 keV, assuming a photon index of 2.0. This assumption was made because the distribution of photon indices in the power law model fitting for 2RXS objects peaks at 2.0 \citep{2016A&A...588A.103B}. The flux error values incorporate the uncertainties from the spectral model fitting.

\subsection{EBXs in Chandra} \label{sec:sec:Sample_selection_Chandra}

We utilized the Chandra Source Catalog 2.0 Quick Search\footnote{http://cda.cfa.harvard.edu/cscweb/index.do} \citep{2019HEAD...1711401E, 2020AAS...23515405E} to look for X-ray counterparts, resulting in 24 sources with a matching radius of $1^{\prime\prime}$, after visual screening of the ASAS-SN light curves. Their X-ray fluxes and uncertainties in the 0.5-7.0 keV band were derived under the power law model with a fixed photon index 2.0 and the Galactic $N_{\rm H}$ in the direction of each source, obtained from the $Colden$ tool\footnote{https://cxc.harvard.edu/toolkit/colden.jsp}. Using the $PIMMS$, we converted the unabsorbed X-ray flux in the $0.5-7.0$ keV band into that in the $0.2-12.0$ keV band. The flux error was also calculated, including the uncertainties caused by the assumed underlying spectral model.

\subsection{The full sample size and X-ray source matching background} \label{subsec:X_ray_background}

When combining the samples from different X-ray missions, we adopted the average flux for duplicate sources in the three catalogs. The total number of sources is 255, which constitutes our Full Sample. We adopted $Gaia$ DR3 distances to calculate the X-ray luminosity, along with corresponding uncertainties for each object. The ASAS-SN names, common names, $Gaia$ DR3 IDs, J2000 coordinates (R.A. \& Dec.), orbital periods ($P$), distances, and X-ray luminosity ($\log L_{\rm X}$) with the lower and upper errors are presented in Table~\ref{table:Properties_EBXs}.

\begin{table*}
\caption{Properties of EBXs}
  \begin{center}
  \begin{tabular}{clll}\hline \hline
Index & Column & Units & Description  \\
1  &   ASAS-SN name  &    &              Object name from ASAS-SN catalog \\
2  &   Common names  &    &             Common names \\
3  &   $Gaia$ DR3 ID  &  &                 Unique source identifier from Gaia DR3 \\
4  &   R.A.   &  deg  &                     Right ascension in decimal degrees (J2000) \\
5  &   DEC.   &  deg  &                     Declination in decimal degrees (J2000)  \\
6  &   Period  &  days  &                    Orbital period of the binary system \\
7  &   Distance   &  pc  &                  Linear distance \\
8  &   $A_{\rm G}$     &  mag  &                      Line-of-sight extinction in the G band \\
9  &   $T_{\rm eff}$     &  K  &                    Effective temperature  \\
10  &   $T_{\rm eff}$  lower error   &  K  &          Lower error of Effective temperature \\
11  &   $T_{\rm eff}$  upper error  & K   &          Upper error of Effective temperature \\
12  &   $\log g$        &  dex  &                Surface gravity  \\
13 &   $\log g$  lower error  &  dex  &          Lower error of Surface gravity  \\
14  &   $\log g$  upper error   & dex   &         Upper error of Surface gravity  \\
15  &   $[\rm Fe/\rm H]$      &  dex  &              Metallicity  \\
16  &   $[\rm Fe/\rm H]$ lower error  &  dex  &        Lower error of Metallicity  \\
17  &   $[\rm Fe/\rm H]$ upper error   &  dex  &       Upper error of Metallicity  \\
18  &   $\log L_{\rm X}$            &  erg/s  &           X-ray luminosity in 0.2-12 keV band \\
19  &   $\log L_{\rm X}$  lower error   & erg/s    &        Lower error of $\log L_{\rm X}$  \\
20  &   $\log L_{\rm X}$  upper error   &  erg/s  &        Upper error of $\log L_{\rm X}$  \\
21  &   $\log L_{\rm bol}$              &  erg/s  &        Bolometric luminosity  \\
22  &   $\log L_{\rm bol}$ lower error &  erg/s  &        Lower error of $\log L_{\rm bol}$  \\
23  &   $\log L_{\rm bol}$ upper error  &  erg/s  &       Upper error of $\log L_{\rm bol}$  \\
24  &   $\log(L_{\rm X}/L_{\rm bol})$       &    &         Ratio of X-ray luminosity to bolometric luminosity  \\
25  &   $\log(L_{\rm X}/L_{\rm bol})$ lower error&    &    Lower error of $\log(L_{\rm X}/L_{\rm bol})$  \\
25  &   $\log(L_{\rm X}/L_{\rm bol})$ upper error &    &   Upper error of $\log(L_{\rm X}/L_{\rm bol})$  \\
27  &   V-Ks        &   mag &             The color for V-band magnitude minus Ks-band magnitude \\ 
28  &   $\tau_{t}$     &  days  &         Convective turnover time \\
29  &   $R_{\rm O}$      &    &        Rossby number \\
\hline\noalign{\smallskip}
 \end{tabular}
 \begin{tablenotes}
      \footnotesize
      \item[1] (This table is available in its entirety in the online machine-readable form.)
\end{tablenotes}
 \end{center}
 \label{table:Properties_EBXs}
\end{table*}

We estimated the expected ``background" random match to the X-ray sources using the package $astropy.coordinates$. Firstly, we add a 1$^{\circ}$ offset to each object in a random direction. Then, we employ the same cross-matching method as described in Sections~\ref{sec:Sample_selection_XMM}, \ref{sec:sec:Sample_selection_RASS} and \ref{sec:sec:Sample_selection_Chandra}. The random ``background" rate of X-ray matching is 1.18\% (3/255), which is negligible for our analysis.

\subsection{Effective Temperature, Gravity and Metallicity}\label{sec:Spectra_of_LAMOST}

To further obtain the stellar atmospheric parameters and bolometric luminosity for EBXs, we cross-matched our Full Sample with the Large Sky Area Multi-Object Fiber Spectroscopic Telescope Data Release 9 catalog (LAMOST DR9; \citealt{2012RAA....12..723Z}; 48 counterparts), and the $Gaia$ DR3 catalog \citep{2023A&A...674A...1G}. LAMOST DR9 provides one set of the atmospheric parameters $T_{\rm eff}$, $\log g$, and [Fe/H], while the $Gaia$ DR3 Astrophysical Parameters Supplement Catalog\footnote{The ``I/355/paramsup" catalog on https://vizier.cds.unistra.fr/viz-bin/VizieR} in \citet{2023A&A...674A...1G} provided two sets of $T_{\rm eff}$, $\log g$ and [M/H],\footnote{We treat the [M/H] as [Fe/H] \citep{2014EAS....65...17C}.} each from GSP-Phot Aeneas, for the MARCS (named $Gaia3M$; 190 counterparts) and PHOENIX (named $Gaia3P$; 138 counterparts) libraries, respectively, using BP/RP spectra. The parameter values from the LAMOST DR9, $Gaia3M$, and $Gaia3P$ catalogs are generally consistent with each other. We finally chose the $T_{\rm eff}$, [M/H] and $\log g$ values from $Gaia3M$ because it provides the largest number of counterparts to our Full Sample, while choosing the single optimal catalog can avoid the heterogeneity by combining multiple catalogs. The adopted stellar parameter values are listed in columns 9, 12, and 15 of Table~\ref{table:Properties_EBXs}.

We calculated the bolometric luminosity for the 190 counterparts in the $Gaia3M$ catalog, following the method provided by the $Gaia$ data release documentation\footnote{https://gea.esac.esa.int/archive/documentation/GDR3/} as follows: 
\begin{equation}\label{equ:bolometricL}
    -2.5~\textrm{log}_{10}~(L_{\textrm{bol}}/L_{\odot}) = M_{\textrm{G}} + BC - M_{\rm sun},
\end{equation}
\begin{equation}\label{equ:Mg}
    M_{\textrm{G}} = G + 5 - 5\log_{10} D - A_{\rm G},
\end{equation}
where $M_{\textrm{G}}$ and $G$ are the absolute and apparent $G-$band magnitudes, while $M_{\rm sun}$ is the solar bolometric magnitude of 4.74~mag, $BC$ is the bolometric correction, and $D$ is the distance. $A_{\rm G}$ is the $StarHorse$ extinction in the G-band provided by \citet{2022A&A...658A..91A}. $BC$ is calculated using the \verb+Python+ code\footnote{https://gitlab.oca.eu/ordenovic/gaiadr3\_bcg} provided by \citet{2023A&A...674A..26C} based on the effective temperatures $T_{\rm eff}$, surface gravity $\log g$, iron abundance [Fe/H], and alpha enhancement [$\alpha$/Fe] (which was set to 0). The bolometric luminosity values are listed in column 21 of Table~\ref{table:Properties_EBXs}.

\subsection{Mass and Radius}\label{sec:Mass_&_Radius_&_age}

We developed the single- and binary-star spectral models with machine learning (Liu et al. 2024, in preparation) to facilitate the spectral fitting for the LAMOST DR9 data. We utilized the spectra from LAMOST DR9 and the stellar parameters obtained from the Apache Point Observatory Galactic Evolution Experiment (APOGEE) DR16 as our training dataset. To create the single-star spectral model, we employed the neural network version of the Stellar Label Machine (SLAM) \citep{2020ApJS..246....9Z, 2020RAA....20...51Z}. Subsequently, the binary-star model was constructed by combining two single-star models while considering their respective radial velocities. In our model, we also used the MIST model trained from stellar evolutionary tracks \citep{2016ApJS..222....8D, 2016ApJ...823..102C} to convert mass, age, and metallicity to effective temperature, gravity, and radius. Our single- and binary-star spectral models have been applied in the spectral fitting in the search for compact objects \citep{2023arXiv230803255Z}.

\begin{figure*}
\centering
\includegraphics[width = 15cm]{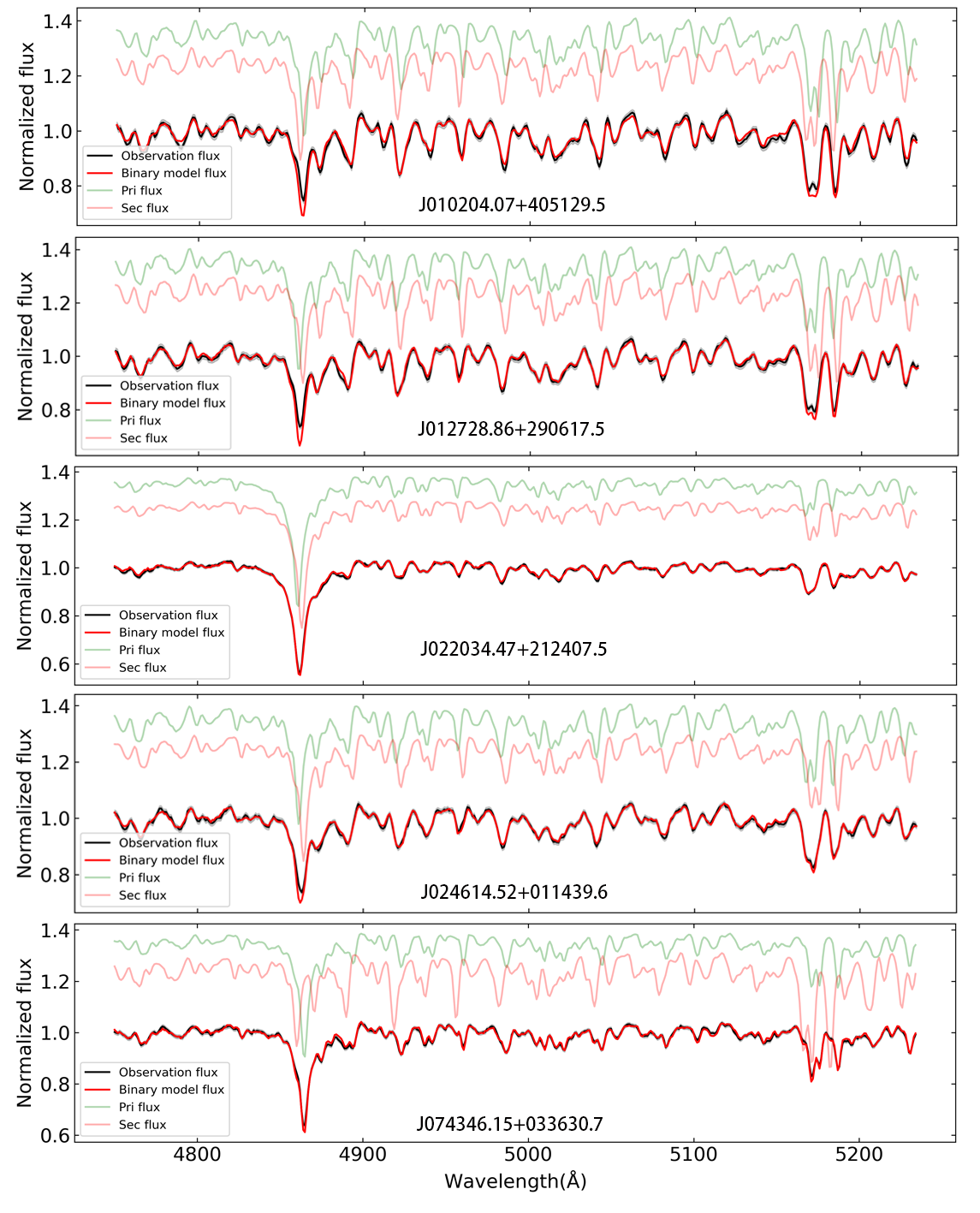}
\caption{The comparison between observed and binary-star model fitting spectra, as well as the corresponding primary and secondary fitting spectra, for five EBXs. The gray area represents the observational errors in the spectra.}
\label{fig:spectral_fitting}
\end{figure*}

We directly performed binary-star model fitting on the LAMOST DR9 spectra of 40 EBXs with the signal-to-noise ratio $S/N > 30$ (out of a total of 48 sources as mentioned in Section~\ref{sec:Spectra_of_LAMOST}). As an example, in Figure~\ref{fig:spectral_fitting}, we present the spectral fitting results for five sources, indicating that the observational spectra can be well-fitted by our model. Eventually, the fitting yielded parameters of each component, i.e., the mass $M$ and radius $R$, are presented in Table~\ref{table:M_R_EBXs}. We use subscripts 1 and 2 to denote the more massive primary and less massive secondary stars, respectively.

We did not apply the temperature, metallicity, and surface gravity derived simultaneously from the binary-star model fitting to Sections~\ref{sec:Spectra_of_LAMOST}, which reduces the potential impact on the statistical results due to the differences in parameter derivation methods. Furthermore, the analysis of the relationship between mass and radius and the properties of X-ray radiation can serve as an independent validation for the analysis of other parameters (e.g., period and effective temperature), as they are obtained through mutually independent methods.

\startlongtable
\begin{deluxetable*}{crrrrrr}
\tablecaption{The Masses and Radii of the components of 40 EBXs \label{table:M_R_EBXs}}
%\tablewidth{500pt}
\tabletypesize{\scriptsize} 
\tablehead{
 ASAS-SN Name & \colhead{R.A.} & \colhead{Dec.} & \colhead{$M_{1}$} & 
\colhead{$M_{2}$} & \colhead{$R_{1}$} &
\colhead{$R_{2}$} \\
\colhead{(ASASSN-V)} & \colhead{(J2000; $^{\circ}$)} & \colhead{(J2000; $^{\circ}$)} & \colhead{($M_{\rm \odot}$)} & 
\colhead{($M_{\rm \odot}$)} & \colhead{($R_{\rm \odot}$)} &
\colhead{($R_{\rm \odot}$)}  
}
\startdata
J012728.86+290617.5 & 21.87026 & 29.10485& $0.960_{-0.006}^{+0.006}$ & $0.877_{-0.009}^{+0.011}$& $1.074_{-0.006}^{+0.006}$& $1.074_{-0.004}^{+0.004}$\\
J005417.76+394510.0 & 13.57401 & 39.75278& $0.933_{-0.006}^{+0.005}$& $0.671_{-0.009}^{+0.013}$& $1.007_{-0.010}^ {+0.012}$& $0.657_{-0.010}^{+0.011}$\\
...  & ...  &... & ... & ... & ... & ... \\
J051609.09+245814.1&  79.03788&  24.97057 & $1.442_{-0.026}^{+0.019}$& $1.051_{-0.029}^{+0.021}$& $1.345_{-0.013}^{+0.012}$& $0.917_{-0.022}^{+0.021}$ \\
J082226.21+205859.5& 125.60922&  20.98319& $1.506_{-0.013}^{+0.017}$& $1.228_{-0.016}^{+0.021}$ & $1.863_{-0.021}^{+0.021}$ & $1.285_{-0.032}^{+0.031}$ \\
\enddata
\begin{tablenotes}
      \footnotesize
      \item[1] (This table is available in its entirety in the online machine-readable form.)
\end{tablenotes}
\end{deluxetable*}

\subsection{Rossby numbers}\label{subsec:Rossby_number}

The Rossby number $R_{\rm O}$, first defined by \citet{1984ApJ...287..769N}, is a dimensionless quantity used to describe the stellar rotation-activity relation, specifically in the context of stellar activity and the dynamo processes. It is defined as the ratio of the rotation period ($P$) of a star to its convective turnover time $\tau_{t}$, i.e., $R_{\rm O}$=$P$/$\tau_{t}$, which represents the characteristic timescale for convective motions in the star's interior. For the calculation of $\tau_{t}$, we used the empirical color-log($\tau$) relation developed by \citet{2018MNRAS.479.2351W}, which is applicable in the range $1.1 < V-Ks < 7.0$, where $V$ and $Ks$ are the average magnitudes collected from ASAS-SN and 2MASS surveys, respectively. The $A_{\rm V}$ and $A_{\rm Ks}$ (= 0.596$\times A_{\rm V}$; \citealt{2019ApJ...877..116W}) from $Starhorse$ \citep{2022A&A...658A..91A} was used to apply the extinction correction for the observed average $V$-band and $Ks$-band magnitudes, respectively. The values of color $V-Ks$ and $\tau_{t}$ are listed in columns 27 and 28 of Table~\ref{table:Properties_EBXs}, while the Rossby number $R_{\rm O}$ values are listed in the last column.

\section{Data Analysis} \label{sec:data_analysis}

\subsection{Distribution in $\log T_{\rm eff}-\log g$ space} \label{subsec:Teff-logg_space}

\begin{figure*}
\centering
\includegraphics[trim=10 0 30 0, clip, width = 12cm]{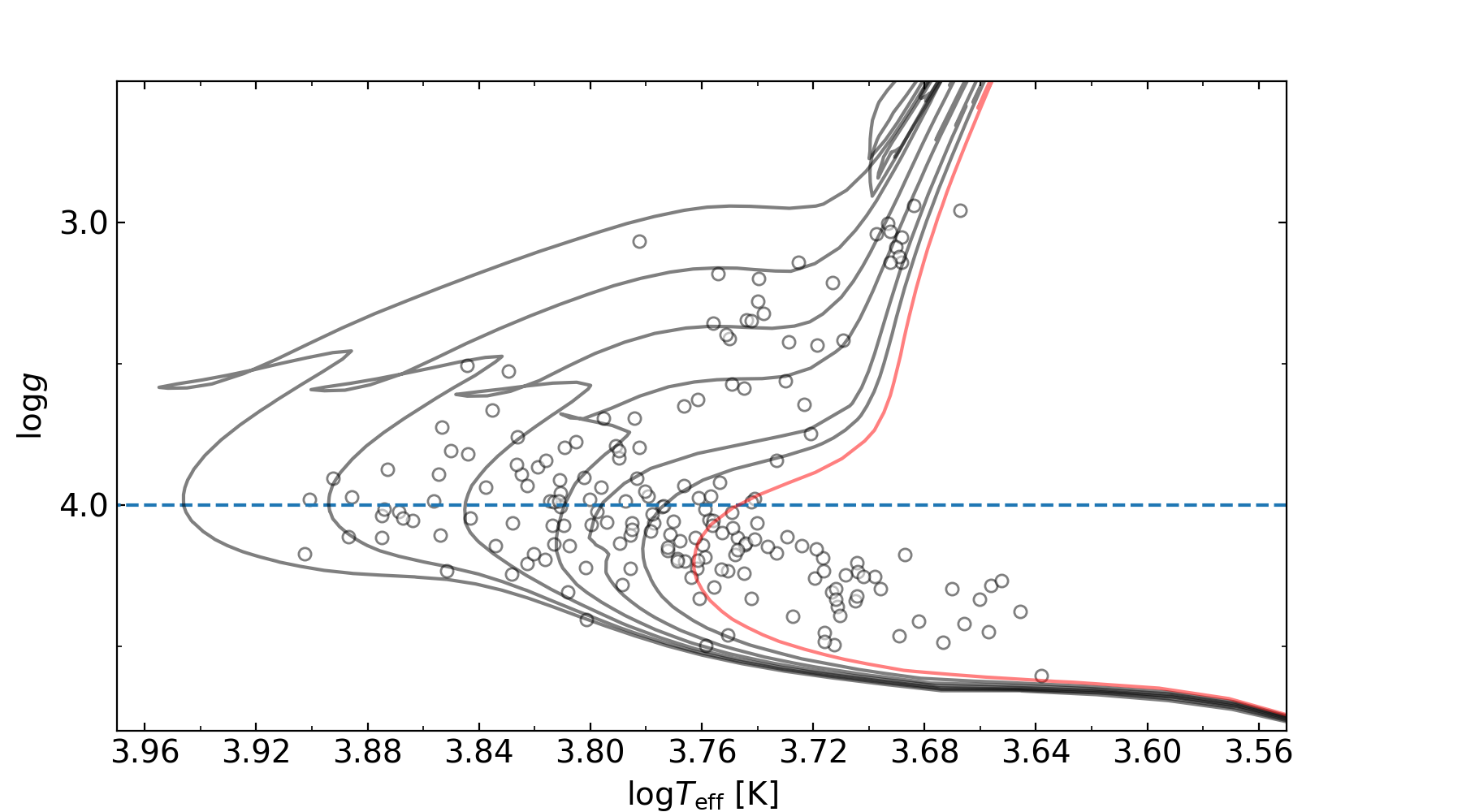}
\caption{The distribution in $\log T_{\rm eff}-\log g$ space for EBXs. The blue dashed line is located at $\log g=4.0$. Data points are indicated by circles. The curves are the isochrones with different ages from left to right, ranging from $10^{8.8}$ to $10^{10.0}$ (in red) years in steps of 0.2~dex.}
\label{fig:Teff-logg_space_distributions}
\end{figure*}

The 190 objects in our Full Sample with stellar parameters are used in the subsequent analyses. We plot these objects on the $\log T_{\rm eff}-\log g$ diagram in Figure~\ref{fig:Teff-logg_space_distributions}, where each solid line shows the theoretical isochrone for stars with the same age and different masses. These isochrones, in the range of $10^{8.8}$ to $10^{10.0}$ years with intervals of 0.2 dex, are derived from stellar evolutionary tracks computed with PARSEC \citep[version 1.2S,][]{2012MNRAS.427..127B}, using solar metallicity. The sample is divided into two sub-samples based on the values of $\log g$: $Sample~1$ with $\log g>4.0$ and $Sample~2$ with $\log g<4.0$. $Sample~1$ contains objects where surface gravity generally decreases with increasing temperature, indicating that the radii of these stars increase with increasing temperature. In contrast, $Sample~2$ have objects whose surface gravity increases with increasing temperature, suggesting that their radii decrease with increasing temperature. We can infer that $Sample~1$ mainly consists of main-sequence stars, while $Sample~2$ is mainly composed of sub-giants and giants, as well as a portion of stars about to depart from the main sequence. The investigations in $\log L_{\rm X}$ and $\log(L_{\rm X}/L_{\rm bol})$ in the following sections also show that they have different X-ray emission properties. It is worth noting that the determination of spectral parameters ($T_{\rm eff}$, $\log g$, and [Fe/H]) did not consider the influence of binaries. \citet{2018MNRAS.473.5043E} pointed out that the temperature difference when fitting binaries with a single-star model in LAMOST is around 100~K, and the difference in $\log g$ is about 0.1~dex, both of which are much smaller than the overall parameter distribution range (4300~K$<T_{\rm eff}<$7900~K; 2.90$<\log g<$4.60). We do not expect a significant impact of these uncertainties on the analysis presented in this work.

% for the former, their surface gravity decreases with increasing temperature, indicating that the radii of these stars increase with increasing temperature, suggesting that they mainly correspond to main sequence stars. As for the latter, generally, the surface gravity of these stars increases with decreasing temperature, suggesting that their radii increase with decreasing temperature. This means they are about to leave the main sequence and evolve into sub-giants and giants.

\subsection{X-ray emission versus period} \label{subsec:X-ray_P}

\begin{figure*}
\centering
\includegraphics[width = 12cm]{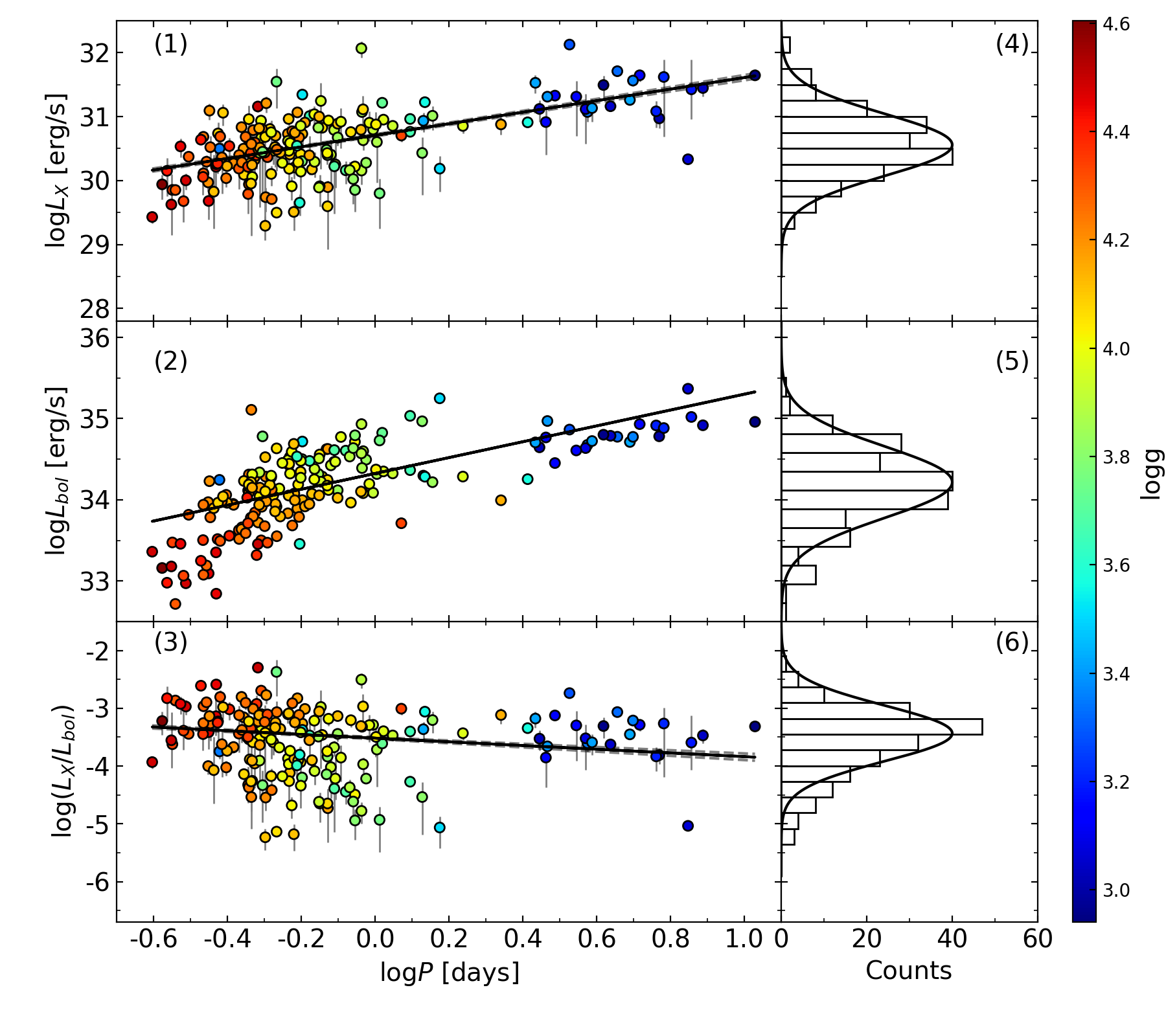}
\caption{The relationships between orbital period $\log P$ and X-ray luminosity $\log L_{\rm X}$, bolometric luminosity $\log L_{\rm bol}$, and X-ray activity level $\log(L_{\rm X}/L_{\rm bol})$. In the middle panel, the error bars of $\log L_{\rm bol}$ are smaller than the size of the symbols. The color map on the right illustrates the surface gravity values.}
\label{fig:Lx_Lbol_distributions}
\end{figure*}

We investigate the correlation between the orbital period and the X-ray emission for our EBXs. We performed linear regression for the correlation analysis between $\log P$ versus $\log L_{\rm X}$, $\log L_{\rm bol}$ and $\log(L_{\rm X}/L_{\rm bol})$ using the Markov Chain Monte Carlo (MCMC) fitting procedure (see Figure~\ref{fig:Lx_Lbol_distributions}). The results are as follows:
\begin{equation}\label{equ:EBX_LX_logP}
    \log L_{\rm X} = (0.90 \pm 0.01) \times \log P + (30.71 \pm 0.01),
\end{equation}
\begin{equation}\label{equ:EBX_Lbol_logP}
    \log L_{\rm bol} = (0.98\pm 0.01) \times \log P + (34.32 \pm 0.01),
\end{equation}
\begin{equation}\label{equ:EBX_LX/Lbol_logP}
    \log (L_{\rm X}/L_{\rm bol}) = (-0.32 \pm 0.01) \times \log P - (3.52 \pm 0.01).
\end{equation}
The $\tau$ parameter of Kendall$^{\prime}$s $\tau$ test \citep{1990Rank} for these three relationships is 0.38, 0.56, and -0.11, respectively. The former two relationships have $1-P_{\tau} >99.99\%$ ($>5\sigma$) while the last has $1-P_{\tau} = 98.08\%$ ($\sim 3\sigma$), where $P_{\tau}$ is the null-hypothesis probability.

The statistical distributions of $\log L_{\rm X}$, $\log L_{\rm bol}$ and $\log(L_{\rm X}/L_{\rm bol})$ are shown in the right panels of Figure~\ref{fig:Lx_Lbol_distributions}. They generally follow the normal distributions with the best-fit parameters of $\mu = 30.56\pm0.02$, 34.22$\pm0.05$ and -3.43$\pm0.04$, and $\sigma = 0.48\pm0.02$, 0.44$\pm0.06$ and 0.48$\pm0.04$, respectively. The $\log L_{\rm X}$ (erg/s) range from 29.3 to 32.1, while the $\log(L_{\rm X}/L_{\rm bol})$ ranges from -5.2 to -2.3.

It is evident that the surface gravity is negatively correlated with the period (see the color map in Figure~\ref{fig:Lx_Lbol_distributions}), which is not surprising given that lower surface gravity stars tend to have larger radii and thus longer orbital periods of the binary system (also see Section~\ref{subsection:magnetic_R}).

\subsection{Stellar Parameters Analysis}\label{sec:Spectral_parameters}

\subsubsection{Effective Temperature}\label{subsubsec:Teff}

\begin{figure*}
\centering
\includegraphics[trim = 0 10 10 10 ,width = 17cm]{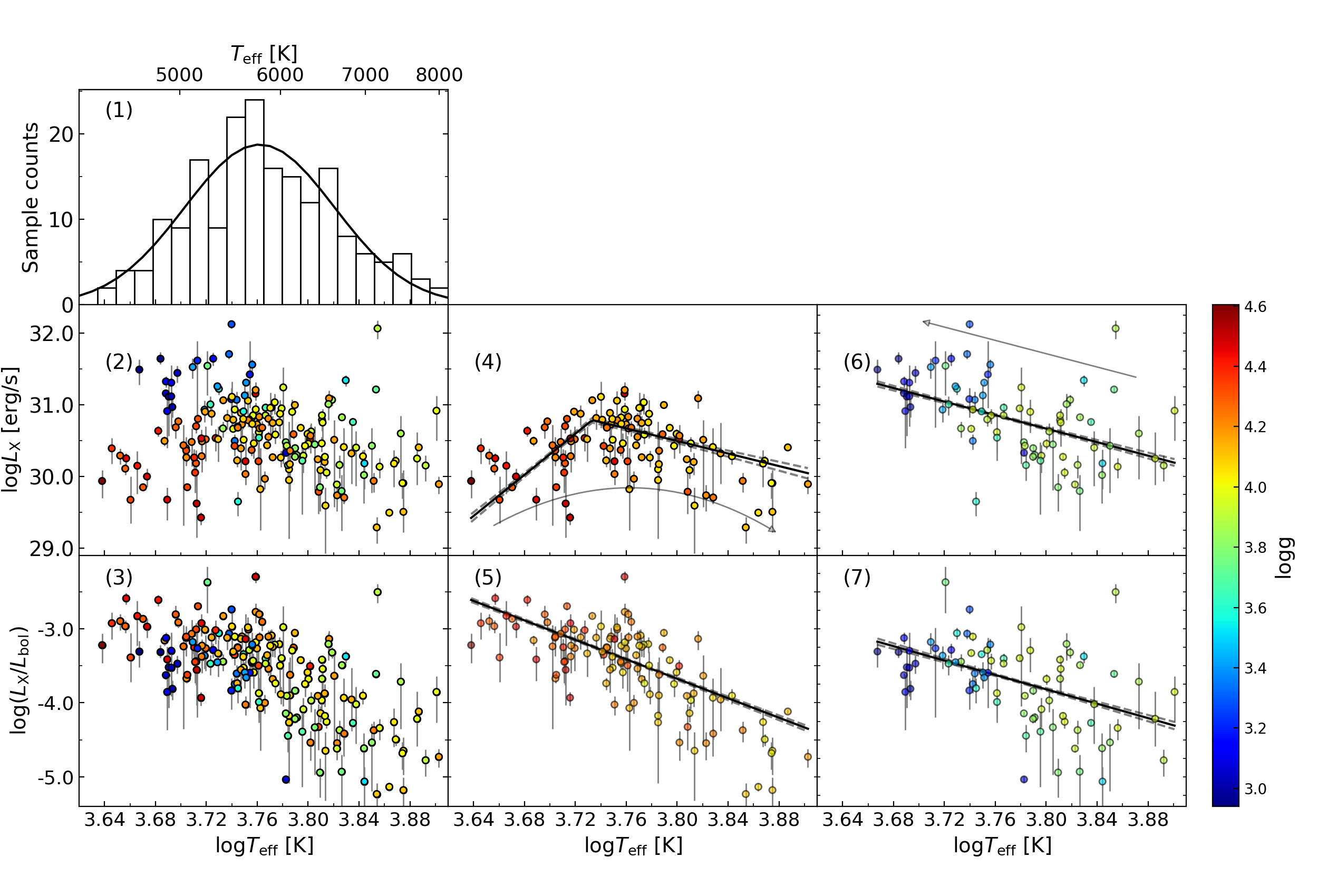}
\caption{The distribution of $T_{\rm eff}$ of our sample is plotted in panel (1) with the best-fit Gaussian profile. Panels (2) and (3) illustrates the $\log T_{\rm eff}$-$\log L_{\textrm{X}}$ and $\log T_{\rm eff}$-$\log (L_{\textrm{X}}/L_{\textrm{bol}})$ relationships, respectively, with the symbols color-coded by $\log g$ (see the right colorbar).
Panels (4) and (6) are the distributions for $Sample~1$ and $Sample~2$ in $\log T_{\rm eff}$-$\log L_{\textrm{X}}$ relation, respectively. Panels (5) and (7) are the distributions for $Sample~1$ and $Sample~2$ in $\log T_{\rm eff}$-$\log (L_{\textrm{X}}/L_{\textrm{bol}})$ relation, respectively.
The black lines are the (segmented) linear fitting results. The dashed lines in all panels represent 95\% uncertainty ranges of the MCMC fitting. The arrows show the direction of the decreasing $\log g$.}
\label{fig:Teff_Lx_Lbol_distributions}
\end{figure*}

In Figure~\ref{fig:Teff_Lx_Lbol_distributions}, panel (1) shows the distribution of effective temperatures of EBXs in Gaussian fitting with $\mu=3.76\pm0.04$ ($\sim 5700$ K) and $\sigma=0.06\pm0.04$. The $\log T_{\rm eff}$ versus $\log L_{\rm X}$ and $\log(L_{\rm X}/L_{\rm bol})$ relationships are plotted with small circles in panels (2) and (3), respectively. 
Based on the classification in Section~\ref{subsec:Teff-logg_space}, we separately plotted $Sample~1$ and $Sample~2$ in panels (4) and (6). 
It is evident that there is a ``turning" point at $\log T_{\rm eff} \sim 3.73$ in the $\log L_{\rm X}$-$\log T_{\rm eff}$ relation for $Sample~1$, while for $Sample~2$, $\log L_{\rm X}$ is anti-correlated with $\log T_{\rm eff}$. The thin arrows in panels (4) and (6) indicate the directions of decreasing $\log g$. As defined in Section~\ref{subsec:Teff-logg_space}, objects in $Sample~2$ have smaller $\log g$ values than those in $Sample~1$. It is worth noting that $Sample~2$ objects generally have higher X-ray luminosity than $Sample~1$ objects. 

We used the segmented linear function to fit this distribution, and the marginalized posterior probability distributions are shown in Figure~\ref{fig:Teff_Lx_Lbol_corner} in the Appendix. The fitting result is listed as follows, 
\begin{equation*}
\begin{split}
\label{equ:Teff_Lx}
\log L_{\rm X}=\left\{
\begin{aligned}
~ & k_{1} \times \log T_{\rm eff} + b =  \\
~ & (14.35_{-0.27}^{+0.26}) \times \log T_{\rm eff}
-(22.77\pm0.98),~~~~~~ & (\log T_{\rm eff}\leq3.73), \\
~ & k_{2} \times \log T_{\rm eff} + (k_{1} - k_{2}) \times \log T_{\rm eff, break} + b = \\
~ & -(4.38 \pm 0.18) \times \log T_{\rm eff} +(47.05\pm1.00),~~~~~~ & (\log T_{\rm eff}>3.73).
\end{aligned}
\right.
\end{split}
\end{equation*}
where $k_{1}$ and $k_{2}$ are the slopes of two fitting lines, respectively; $\log T_{\rm eff, break}$ is the breakpoint, and $b$ is the intercept of the first part of the segmented linear fitting. The marginalized posterior probability distribution in Figure~\ref{fig:Teff_Lx_Lbol_corner} shows the break point $\log T_{\rm eff, break}=3.73$ ($T_{\rm eff} \sim 5400$K). The $\log L_{\rm X}$-$\log T_{\rm eff}$ anti-correlation of $Sample~2$ is best represented by the following equation,
\begin{equation}\label{equ:logLX_LogTeff}
    \log L_{\rm X} = (-4.71\pm0.11) \times \log T_{\rm eff} + (48.58\pm 0.48).
\end{equation}
The Kendall's $\tau$ test \citep{1990Rank} shows that $\tau$ is -0.42 with confidence $1-P_{\tau} >99.99\%$ ($>5\sigma$).

In contrast to the unusual behavior of the $\log T_{\rm eff}-\log L_{\rm X}$ relation, both $Sample~1$ and $Sample~2$ exhibit high-confidence ($>5 \sigma$) anti-correlations between the X-ray activity level ($\log(L_{\rm X}/L_{\rm bol})$) and effective temperature. The difference lies in that when we consider along the increasing $\log T_{\rm eff}$ (and thus decreasing $\log(L_{\rm X}/L_{\rm bol})$), the surface gravity $\log g$ decreases in $Sample~1$ but increases in $Sample~2$. The results of the linear fitting and Kendall's $\tau$ test are listed in Table~\ref{table:Teff_log_lx/lbol}.

\begin{table*}
   \caption{The parameters of linear fitting ($\log(L_{\rm X}/L_{\rm bol}) = a \times \log T_{\rm eff} + b$) and Kendall's $\tau$ test for $Sample~1$ and $Sample~2$ in the relation $\log T_{\rm eff}-\log(L_{\rm X}/L_{\rm bol})$.}
   \begin{center}
   \begin{tabular}{lccccc}\hline \hline
$\log T_{\rm eff}-\log(L_{\rm X}/L_{\rm bol})$         &  $a$  &   $b$  &  $\tau$ & 1-$P_{\tau}$  &  \\
$Sample~1$ &  -6.57$\pm0.08$  &  21.30$\pm0.28$ & -0.50  & $>$99.99$\%$  \\
$Sample~2$ &   -4.85$\pm0.11$ &  14.65$\pm0.43$ & -0.36  & $>$99.99$\%$  \\
\hline\noalign{\smallskip}
  \end{tabular}
  \end{center}
  \label{table:Teff_log_lx/lbol}
\end{table*}

\subsubsection{Metallicity and Surface Gravity}\label{subsubsec:MH_logg}

\begin{figure*}
\centering
\includegraphics[width = 12cm]{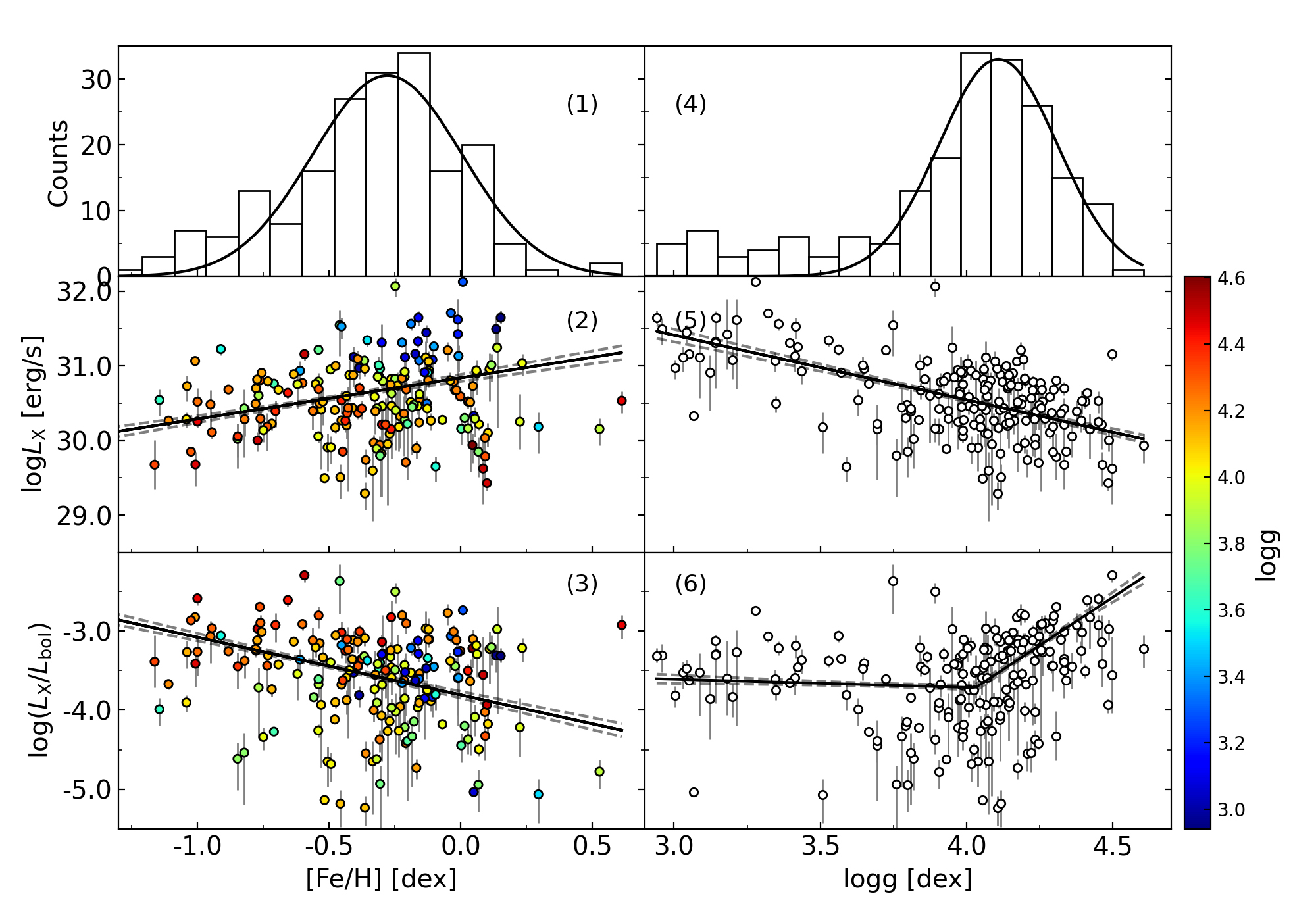}
\caption{The relationships for the metallicity [$\rm Fe/\rm H$] and surface gravity $\log g$ versus X-ray luminosity $\log L_{\textrm{X}}$ and activity level $\log(L_{\textrm{X}}/L_{\textrm{bol}})$. Panels (1) and (4) are the distributions of $[\rm Fe/ \rm H]$ and $\log g$ with Gaussian fittings. The dashed lines in all panels represent 95\% uncertainty ranges of the MCMC fitting.}
\label{fig:feh_logg_Lx_Lbol}
\end{figure*}

The statistical distributions of the metallicity and surface gravity values of our sample are both modeled with Gaussian profiles, resulting in the best-fit parameters of $(\mu, \sigma) = (-0.28\pm0.03, 0.28\pm0.07)$ for [Fe/H], and 
$(\mu, \sigma) = (4.11\pm0.01, 0.20\pm0.02)$ for $\log g$ (see panels 1 \& 4 in Figure~\ref{fig:feh_logg_Lx_Lbol}). The tail towards smaller $\log g$ indicates the giant and sub-giant star population. 

Linear regressions with MCMC and Kendall's $\tau$ test are performed on the $[\rm Fe/\rm H]-\log L_{\textrm{X}}$ , $[\rm Fe/\rm H]-\log(L_{\textrm{X}}/L_{\textrm{bol}})$, and $\log g-\log L_{\textrm{X}}$ relations. The best-fit parameters are listed in Table~\ref{table:feh_logg_vs_logLx_log_lx/lbol}. The metallicity $[\rm Fe/\rm H]$ is marginally correlated with $\log(L_{\textrm{X}}/L_{\textrm{bol}})$ at $<3\sigma$ significance. The surface gravity $\log g$ has strong anti-correlation with $\log L_{\textrm{X}}$ at a confidence level of $>5\sigma$. 
For the $\log g-\log (L_{\textrm{X}}/L_{\textrm{bol}})$ relation, the segmented linear fitting with MCMC was employed and listed in Equation~\ref{equ:logg_LxLbol}. The marginalized posterior probability distributions of this fitting are shown in Figure~\ref{fig:feh_logg_Lx_Lbol_corner}, which indicates that the $\log g-\log(L_{\textrm{X}}/L_{\textrm{bol}})$ relationship has a breakpoint at $\log g_{\rm break}=4.03\pm0.02$. This value is consistent with the dividing value $\log g =4.0$ for $Sample~1$ and $Sample~2$ within $2\sigma$, reinforcing the differences in the X-ray activity level of the two subsamples.

\begin{table*}
   \caption{The parameters of least-square fitting and Kendall's $\tau$ test for the metallicity $[\rm Fe/ \rm H]$ and surface gravity $\log g$ versus X-ray luminosity $\log L_{\textrm{X}}$ and activity level $\log(L_{\textrm{X}}/L_{\textrm{bol}})$.}
   \begin{center}
   \begin{tabular}{lccccc}\hline \hline
Parameters          &  $a$  &   $b$  &  $\tau$ & 1-$P_{\tau}$  &  \\
$[\rm Fe/\rm H]-\log L_{\textrm{X}}$ &  $0.55\pm 0.03$  &  $30.84\pm0.01$ & 0.09  & 93.98$\%$  \\
$[\rm Fe/\rm H]-\log(L_{\textrm{X}}/L_{\textrm{bol}})$ &  $-0.72\pm0.03$  &  $-3.81\pm0.02$ & -0.10  & 96.77$\%$  \\
$\log g-\log L_{\textrm{X}}$ &  $-0.87\pm0.03$  &  $34.01\pm0.01$ & -0.32  & $>$99.99$\%$  \\

\hline\noalign{\smallskip}
  \end{tabular}
  \end{center}
  \label{table:feh_logg_vs_logLx_log_lx/lbol}
\end{table*}

\begin{equation}
\label{equ:logg_LxLbol}
\log L_{\rm X}=\left\{
\begin{aligned}
~ & k_{1} \times \log g + b 
=  \\
~ & -(0.10\pm0.02) \times \log g -(3.30\pm0.07),~~~~~~ & (\log g\leq4.03), \\
~ & k_{2} \times \log g + (k_{1} - k_{2}) \times \log g_{\rm break} + b = \\
~ & (2.43\pm0.07) \times \log g -(13.50\pm0.10),~~~~~~ & (\log g>4.03).
\end{aligned}
\right.
\end{equation}

\subsubsection{Magnetic activity and Stellar Mass}\label{subsubsec:data_analysis_M}

\begin{figure*}
\centering
\includegraphics[width = 19cm]{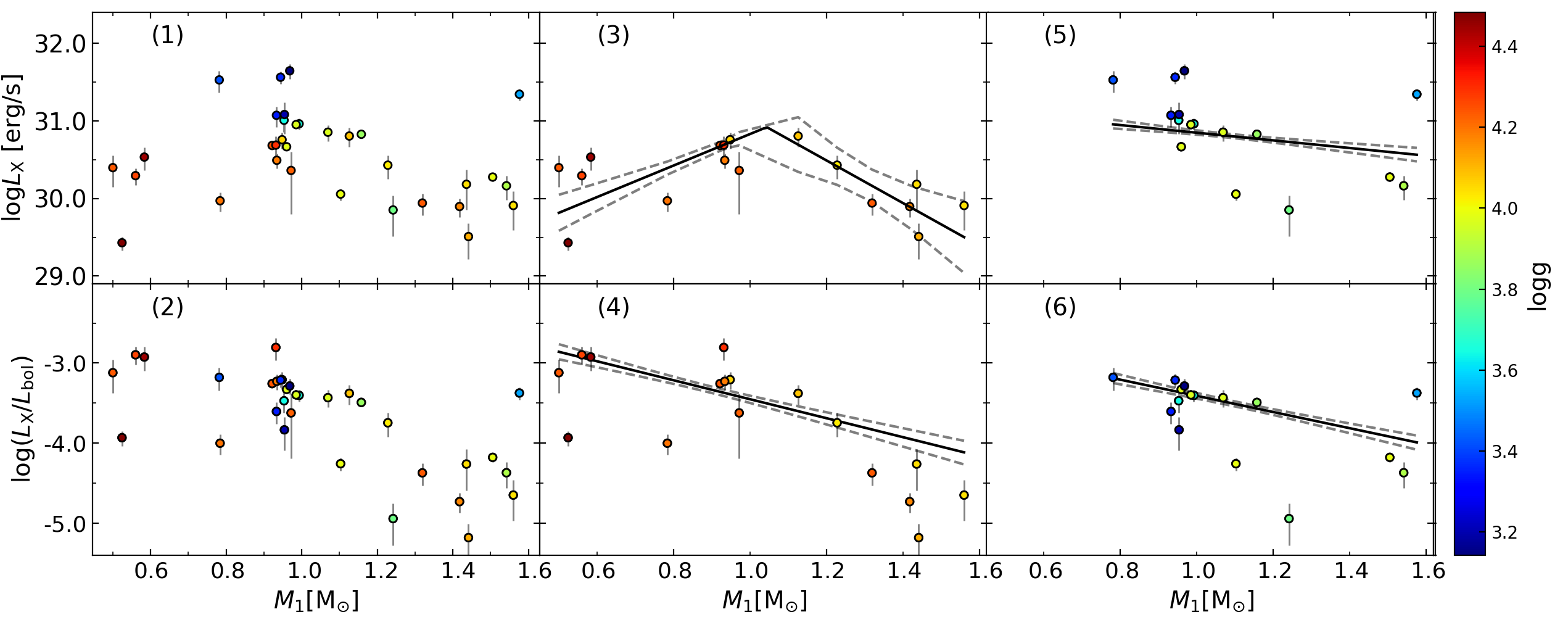}
\caption{The relationships for the primary components masses $M_{1}$ versus X-ray luminosity $\log L_{\textrm{X}}$ (panel 1) and activity level $\log(L_{\textrm{X}}/L_{\textrm{bol}})$ (panel 2). Panels (3) and (4) are for the $Sample~1$, while panels (5) and (6) are for the $Sample~2$. The dashed lines in all panels represent 95\% uncertainty ranges of the MCMC fitting.}
\label{fig:Mass1_vs_Lx_Lbol}
\end{figure*}

Mass, a fundamental stellar parameter, dictates a star's temperature across various evolutionary stages, offering insights into the relationship between temperature and magnetic activity. It should be noted that each binary system corresponds to one X-ray counterpart. We carried out the analysis for the masses of primary component $M_{1}$ versus the binary X-ray luminosity $\log L_{\textrm{X}}$ and the magnetic activity level $\log(L_{\textrm{X}}/L_{\textrm{bol}})$ in Figure~\ref{fig:Mass1_vs_Lx_Lbol}, which is classified based on $Sample~1$ and $Sample~2$. Overall, in panel~(1), there is an increase followed by a decrease in X-ray luminosity in the direction of increasing mass.

In panel (3), the 'peak-like' relationship between $M_{1}$ and $\log L_{\textrm{X}}$ in $Sample~1$ can be well described by the segmented linear model specified in Equation~\ref{equ:M1_Lx_group1} with a peak at $M_{\rm 1, break}=1.04^{+0.03}_{-0.04}$~$\rm M_{\odot}$ (corresponding to a temperature of $\sim5900^{+80}_{-120}$K in the main sequence; \citealt{2000asqu.book.....C}). The marginalized posterior probability distribution is shown in the left panel of Figure~\ref{fig:M1_Lx_Lbol_corner}. For sources in the main-sequence stage with $M_{1}<1.04~\rm M_{\odot}$, their X-ray luminosity increases with the primary star's mass, whereas for sources with $M_{1}>1.04~\rm M_{\odot}$, this trend is the opposite.

In panel (5), for the sub-giants and giants sources in $Sample~2$, their fitted line presented in Equation~\ref{equ:M1_Lx_group2} implies only a decrease in their X-ray luminosity with increasing mass. The Kendall's $\tau$ test \citep{1990Rank} shows that the $\tau$ is -0.46 with confidence $1-P_{\tau} = 98.85\%$ ($>2\sigma$). Moreover, the objects from $Sample~2$ have higher X-ray luminosity compared to those from $Sample~1$ in the mass range of $\sim$0.8 to 1.6~$\rm M_{\odot}$, as indicated by the former consistently being located above the latter at a certain mass. This indicates that the X-ray radiation luminosity of the sub-giants and giants is likely higher than that of the main-sequence stars with similar mass.

\begin{equation}
\begin{split}
\label{equ:M1_Lx_group1}
\log L_{\rm X}=\left\{
\begin{aligned}
(2.03\pm0.19) \times M_{1}
+(28.80\pm0.17),~~~~~~(M_{1}\leq1.04), \\
-(2.75\pm0.53) \times M_{1} +(33.77\pm0.84),~~~~~~(M_{1}>1.04).
\end{aligned}
\right.
\end{split}
\end{equation}

\begin{equation}\label{equ:M1_Lx_group2}
    \log L_{\rm X} = (-0.49\pm0.05) \times M_{1} + (31.34\pm0.06).
\end{equation}

Figure~\ref{fig:Mass1_vs_Lx_Lbol} panel (2) shows a trend of monotonous decrease with primary stars' masses, indicating that as the mass increases, the level of X-ray activity weakens.
In Figure~\ref{fig:Mass1_vs_Lx_Lbol} panels (4) and (6), all the mass-magnetic activity level relationships follow similar negative correlations with high significance ($>2 \sigma$) as listed in Table~\ref{table:M1_log_lx/lbol}, which means that the lower-mass EBX holds higher levels of magnetic activity compared to a higher-mass one.

\begin{table*}
   \caption{The parameters of linear fitting ($\log(L_{\rm X}/L_{\rm bol}) = a \times M_{1} + b$) and Kendall's $\tau$ test for $Sample~1$ and $Sample~2$ in the relation $M_{1}-\log (L_{\rm X}/L_{\rm bol})$.}
   \begin{center}
   \begin{tabular}{lccccc}\hline \hline
$M_{1}-\log(L_{\rm X}/L_{\rm bol})$         &  $a$  &   $b$  &  $\tau$ & 1-$P_{\tau}$  &  \\
$Sample~1$ &  $-1.18\pm0.08$  &  -$2.27\pm0.07$ & -0.56  & 99.87$\%$  \\
$Sample~2$ &  $-1.01\pm0.06$  &   $2.40\pm0.06$ & -0.42  & 97.44$\%$  \\
\hline\noalign{\smallskip}
  \end{tabular}
  \end{center}
  \label{table:M1_log_lx/lbol}
\end{table*}

\begin{figure*}
\centering
\includegraphics[width = 19cm]{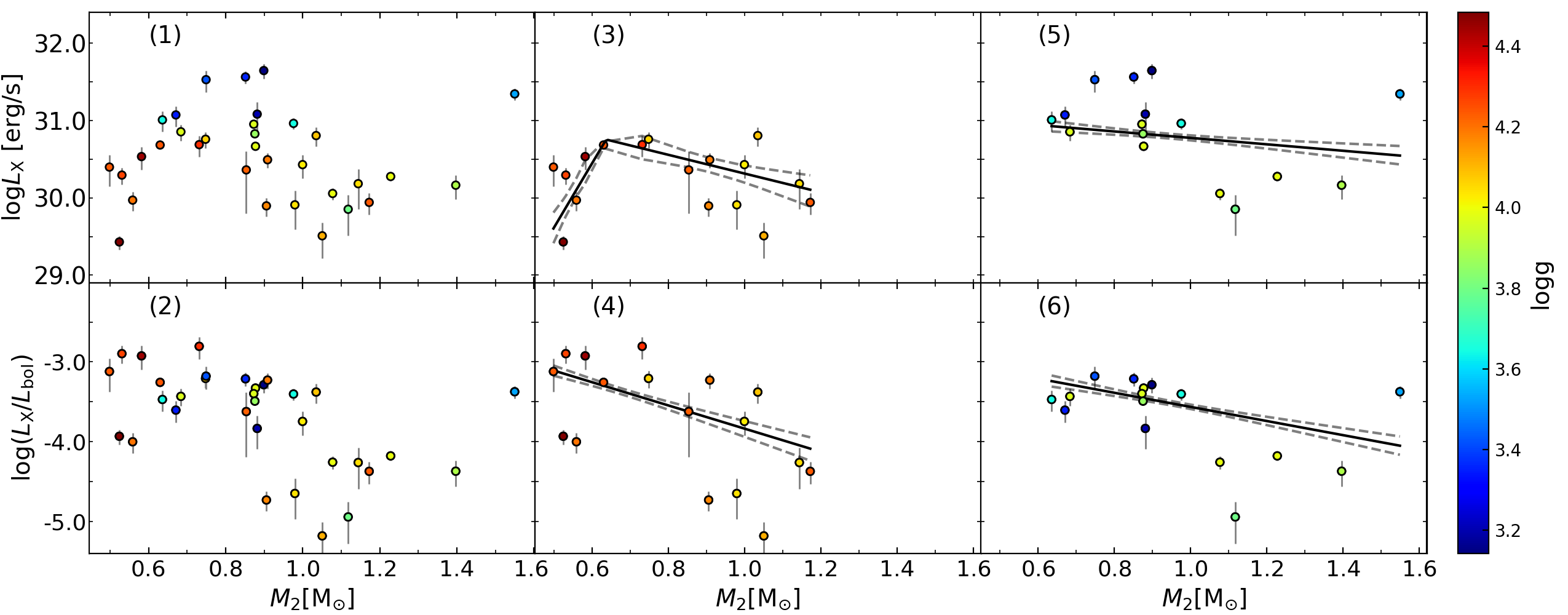}
\caption{The relationships for the secondary components masses $M_{2}$ versus X-ray luminosity $\log L_{\textrm{X}}$ (panel 1) and activity level $\log(L_{\textrm{X}}/L_{\textrm{bol}})$ (panel 2). Panels (3) and (4) are for the $Sample~1$, while panels (5) and (6) are for the $Sample~2$. The dashed lines in all panels represent 95\% uncertainty ranges of the MCMC fitting.}
\label{fig:Mass2_vs_Lx_Lbol}
\end{figure*}

The same analytical processes are also applied to the secondary component in Figure~\ref{fig:Mass2_vs_Lx_Lbol}, revealing that the statistical results of the secondary star's mass versus the binary X-ray luminosity and magnetic activity level are nearly similar to those of the primary star. The fitting result of $M_{2}-\log L_{\textrm{X}}$ for $Sample~1$ with the peak located at $M_{\rm 2, break}=0.64\pm0.02$~$\rm M_{\odot}$ (corresponding to $\sim4310\pm60$~K; \citealt{2000asqu.book.....C}) is listed in Equation~\ref{equ:M2_Lx_group1}, while the marginalized posterior probability distribution is shown in the right panel of Figure~\ref{fig:M1_Lx_Lbol_corner}. The fitting result of $Sample~2$ is presented in Equation~\ref{equ:M2_Lx_group2} with $\tau =-0.27$ and confidence $1-P_{\tau} >83.47\%$ ($>1\sigma$). 
In Table~\ref{table:M2_log_lx/lbol}, we listed the linear fitting of the $M_{2}-\log(L_{\rm X}/L_{\rm bol})$ for $Sample~1$ and $Sample~2$.

\begin{equation}
\begin{split}
\label{equ:M2_Lx_group1}
\log L_{\rm X}=\left\{
\begin{aligned}
(8.31_{-0.84}^{+1.88}) \times M_{2}
+(25.46_{-1.01}^{+0.52}),~~~~~~(M_{2}\leq0.64), \\
-(1.21_{-0.43}^{+0.27}) \times M_{2} +(31.56_{-1.62}^{+2.21}),~~~~~~(M_{2}>0.64).
\end{aligned}
\right.
\end{split}
\end{equation}

\begin{equation}\label{equ:M2_Lx_group2}
    \log L_{\rm X} = (-0.41\pm0.06) \times M_{2} + (31.19\pm0.06).
\end{equation}

\begin{table*}
   \caption{The parameters of linear fitting ($\log(L_{\rm X}/L_{\rm bol}) = a \times M_{2} + b$) and the Kendall's $\tau$ test for $Sample~1$ and $Sample~2$ in the relation $M_{2}-\log(L_{\rm X}/L_{\rm bol})$.}
   \begin{center}
   \begin{tabular}{lccccc}\hline \hline
$M_{2}-\log(L_{\rm X}/L_{\rm bol})$         &  $a$  &   $b$  &  $\tau$ & 1-$P_{\tau}$  &  \\
$Sample~1$ &  $-1.45\pm0.10$  &  -$2.38\pm0.07$ & -0.41  & 97.82$\%$  \\
$Sample~2$ &  $-0.88\pm0.06$  &  -$2.68\pm0.06$ & -0.28  & 86.01$\%$  \\
\hline\noalign{\smallskip}
  \end{tabular}
  \end{center}
  \label{table:M2_log_lx/lbol}
\end{table*}

\subsubsection{Magnetic activity and Stellar Radius}\label{subsection:magnetic_R}

\begin{figure*}
\centering
\includegraphics[width = 19cm]{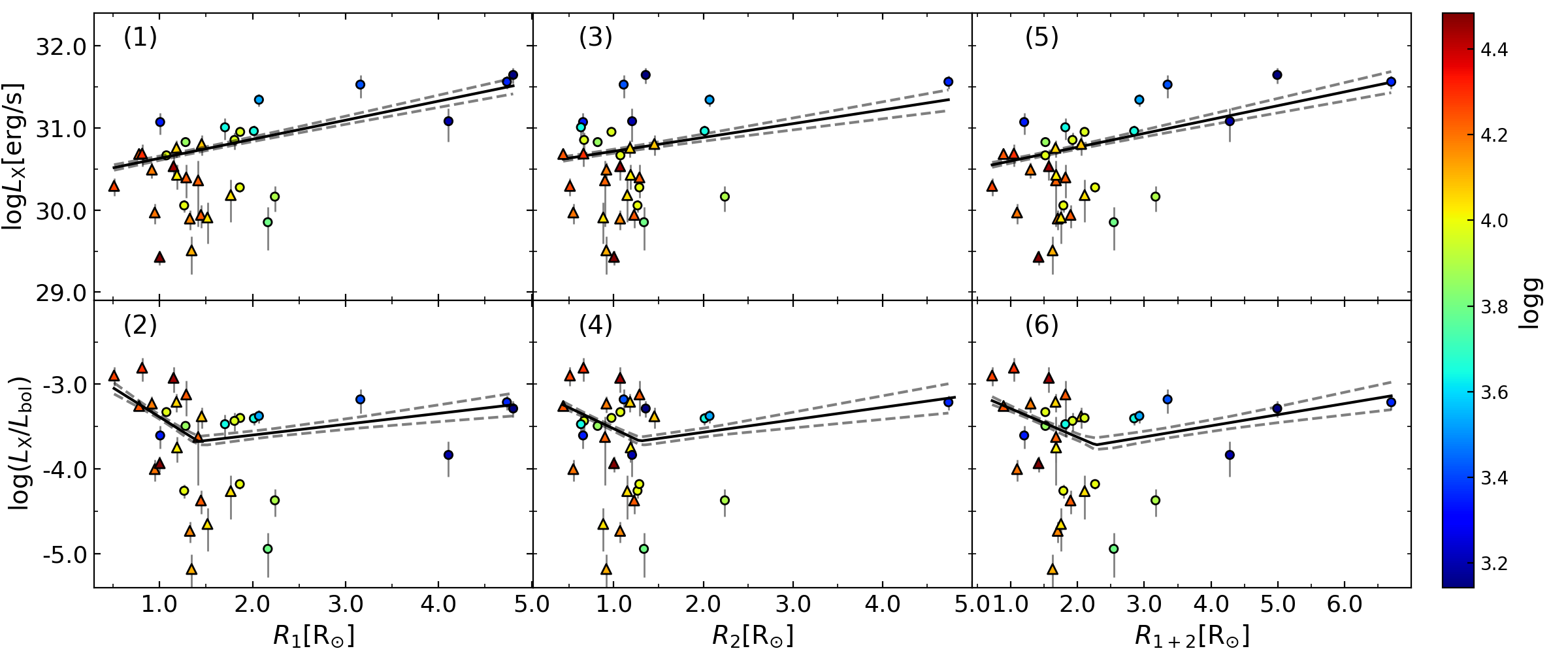}
\caption{The relationships for the radii of primary components $R_{1}$ (panel 1 and 2), secondary components $R_{2}$ (panel 3 and 4) and the binary system equivalent radii $R_{1+2}$ (panel 5 and 6) versus X-ray luminosity $\log L_{\textrm{X}}$ and magnetic activity level $\log(L_{\textrm{X}}/L_{\textrm{bol}})$. The objects from $Sample~1$ and $Sample~2$ are represented by triangles and circles, respectively. The dashed lines in all panels represent 95\% uncertainty ranges of the MCMC fitting.}
\label{fig:radius_vs_Lx_Lbol}
\end{figure*}

The study of radius can directly link the magnetic activity properties with the geometric structure of EBXs. In Figure~\ref{fig:radius_vs_Lx_Lbol}, we investigated the correlation between the radii of primary components $R_{1}$, secondary components $R_{2}$ and binary systems' equivalent radii $R_{1+2}$ ( = $\sqrt{4\times \pi (R_{1}^{2} + R_{2}^{2})/(4\times \pi)}$) of EBXs with the binary X-ray luminosity $\log L_{\textrm{X}}$ and activity level $\log(L_{\textrm{X}}/L_{\textrm{bol}})$. $Sample~1$ and $Sample~2$ are distinguished by triangles and circles, respectively. The fitting results for $R-\log L_{\textrm{X}}$ are listed in Table~\ref{table:R_vs_logLx_log_lx/lbol}. Compared to $R_{2}$ (with confidence $<1\sigma$), the X-ray luminosity shows a high-confidence ($\sim 2\sigma$) positive correlation with $R_{1}$ and $R_{1+2}$. All of the above positive correlations establish that X-ray luminosity is proportional to the radii of EBXs, meaning that X-ray luminosity is proportional to the surface area of EBXs.

For the magnetic activity level, we find that all the $R-\log(L_{\textrm{X}}/L_{\textrm{bol}})$ can be described by the segmented linear fitting (using the same fitting procedure as Equation~\ref{equ:Teff_Lx}) with breaks at $R_{\rm 1, break}=1.42\pm 0.05$~$\rm R_{\odot}$, $R_{\rm 2, break}=1.29^{+0.03}_{-0.02}$~$\rm R_{\odot}$ and $R_{\rm 1+2, break}=2.27^{+0.05}_{-0.03}$~$\rm R_{\odot}$. The fitting results are listed in Equations~\ref{equ:R1_LxLbol}, \ref{equ:R2_LxLbol} and \ref{equ:R12_LxLbol} with the marginalized posterior probability distributions in Figure~\ref{fig:R1_Lx_Lbol_corner}. These results indicate that the magnetic activity level of EBXs first decreases and then increases with the growth of the radius.

Along the direction of increasing radius and decreasing surface gravity, one can find that the distributions of $R-\log L_{\textrm{X}}$ versus $\log g-\log L_{\textrm{X}}$ and $R-\log( L_{\textrm{X}}/L_{\textrm{bol}})$ versus $\log g-\log(L_{\textrm{X}}/L_{\textrm{bol}})$ are consistent. These two consistencies establish the correspondence between the magnetic activity of EBXs at two levels: geometric structure ($R$) and atmospheric parameters ($\log g$), and mutually validate each other.

Moreover, as shown in panels (2), (4) and (6), the clear differences in magnetic activity levels between $Sample~1$ and $Sample~2$ are evident in all $R-\log(L_{\textrm{X}}/L_{\textrm{bol}})$ relationships. The sources in $Sample~1$ (triangle points) mostly follow a negative correlation distribution, while those in $Sample~2$ (circle points) follow a positive correlation distribution. This once again confirms the necessity and correctness of our sample classification (as described in Section~\ref{subsec:Teff-logg_space}) from the perspective of radius. From the color map of surface gravity overlaid on the radius, it's evident that there is a negative correlation between surface gravity and the radii of the component stars of EBXs. All of the above indicates that the components with lower surface gravity tend to have larger radii, confirming the statement made in Sections~\ref{subsec:Teff-logg_space} and \ref{sec:Discu} (the first paragraph): lower surface gravity sources tend to have larger radii, consequently affecting the eclipsing orbital radii and periods.

\begin{equation}
\begin{split}
\label{equ:R1_LxLbol}
\log(L_{\textrm{X}}/L_{\textrm{bol}})=\left\{
\begin{aligned}
(-0.69\pm0.06) \times R_{1}
-(2.70\pm0.06),~~~~~~~(R_{1}\leq1.42), \\
(0.13\pm0.02) \times R_{1} -(3.86\pm0.13),~~~~~~~(R_{1}>1.42).
\end{aligned}
\right.
\end{split}
\end{equation}

\begin{equation}
\begin{split}
\label{equ:R2_LxLbol}
\log(L_{\textrm{X}}/L_{\textrm{bol}})=\left\{
\begin{aligned}
(-0.50\pm0.04) \times R_{2}
-(3.03\pm0.04),~~~~~~~(R_{2}\leq1.29), \\
(0.15\pm0.03) \times R_{2} -(3.87\pm0.10),~~~~~~~(R_{2}>1.29).
\end{aligned}
\right.
\end{split}
\end{equation}

\begin{equation}
\begin{split}
\label{equ:R12_LxLbol}
\log(L_{\textrm{X}}/L_{\textrm{bol}})=\left\{
\begin{aligned}
(-0.34\pm0.03) \times R_{1+2}
-(2.95_{-0.04}^{+0.05}),~~~~~~~(R_{1+2}\leq2.27), \\
(0.13\pm0.02) \times R_{1+2} -(4.02\pm0.10),~~~~~~~(R_{1+2}>2.27).
\end{aligned}
\right.
\end{split}
\end{equation}

\begin{table*}
   \caption{The parameters of least-square fitting and Kendall's $\tau$ test for $R_{1}$, $R_{2}$ and $R_{1+2}$  versus X-ray luminosity $\log L_{\textrm{X}}$.}
   \begin{center}
   \begin{tabular}{lccccc}\hline \hline
Parameters          &  $a$  &   $b$  &  $\tau$ & 1-$P_{\tau}$  &  \\
$R_{1}-\log L_{\textrm{X}}$ &  $0.23\pm0.01$  &  $30.40\pm0.02$ & 0.25  & 96.13$\%$  \\
$R_{2}-\log L_{\textrm{X}}$ &  $0.17\pm0.01$  &  $30.54\pm0.01$ & 0.09  & 51.13$\%$  \\
$R_{1+2}-\log L_{\textrm{X}}$ &  $0.17\pm0.01$  &  $30.43\pm0.02$ & 0.25  & 96.14$\%$  \\
\hline\noalign{\smallskip}
  \end{tabular}
  \end{center}
  \label{table:R_vs_logLx_log_lx/lbol}
\end{table*}

\subsection{Magnetic activity and Rossby number}\label{subsec:Rossby_number_analysis}

\begin{figure*}
\centering
\includegraphics[trim=36 0 30 0, clip, width = 18cm]{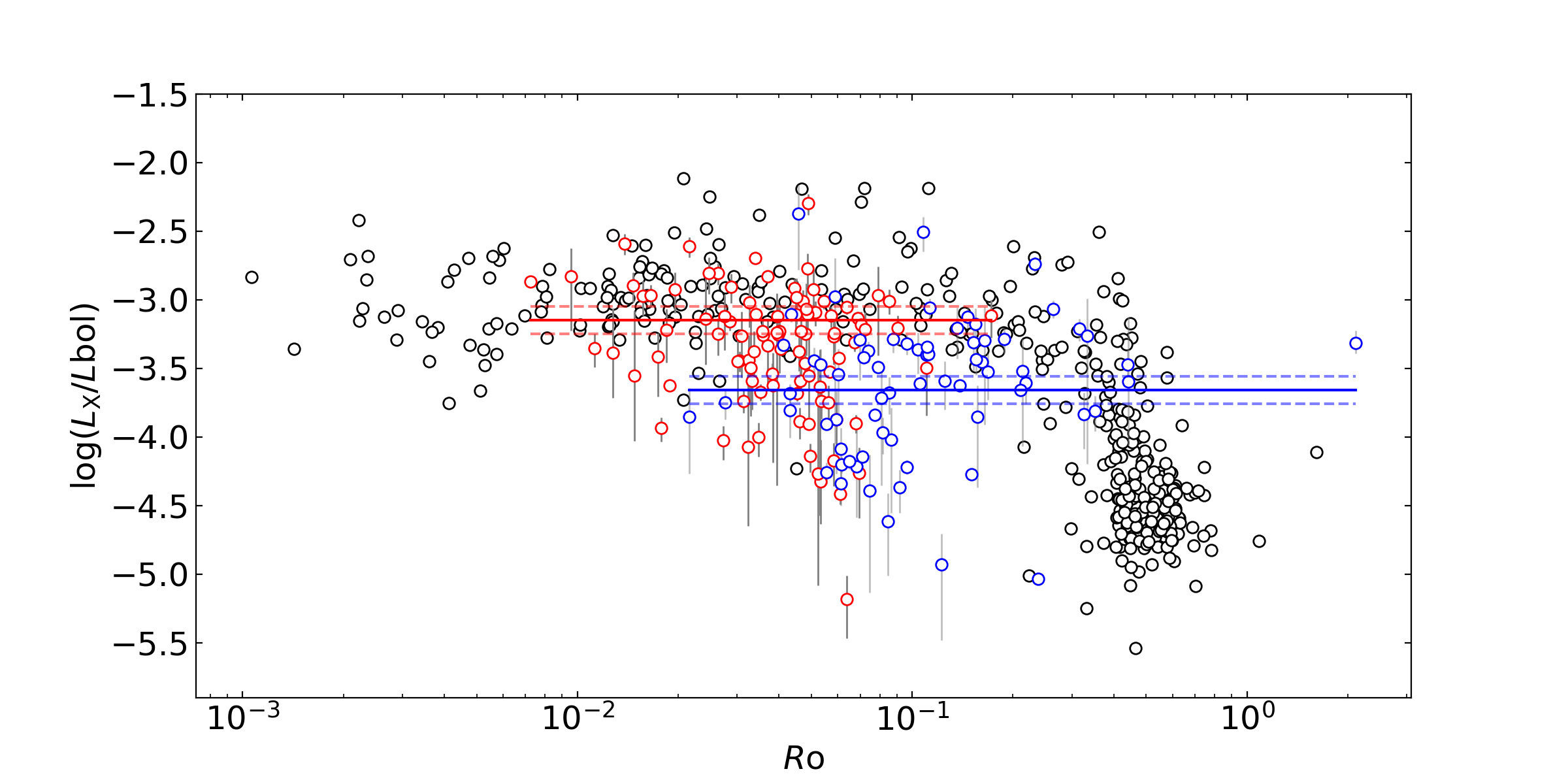}
\caption{$R_{O}$ versus $\log (L_{\rm X}/L_{\rm bol})$ for EBXs. The red and blue open circles are objects from $Sample~1$ and $Sample~2$, respectively, while the red and blue solid lines are the MCMC fitting results, respectively. The dashed lines are the corresponding 1~$\sigma$ fitting error. The black open circles are binary candidates collected from \citet{2022ApJ...931...45N}.}
\label{fig:Ro_analysis}
\end{figure*}

Describing the relationship between $R_{\rm O}$ and $\log (L_{\rm X}/L_{\rm bol})$ provides a more direct way to study coronal activity and rotation in low-mass stars \citep{2022ApJ...931...45N}.
For fitting the $R_{\rm O}$-$\log (L_{\rm X}/L_{\rm bol})$ relation, a widely used model is a constant region connected to a power-law model (e.g., \citealt{2018MNRAS.479.2351W, 2022ApJ...931...45N}), which is shown as follows
\begin{equation}
\label{equ:Rossby_2_parts_equation}
\dfrac{L_{\rm X}}{L_{\rm bol}}=\left\{
\begin{aligned}
(\dfrac{L_{\rm X}}{L_{\rm bol}})_{sat},~~~~~~({\rm if}~R_{\rm O}\leq R_{\rm O,sat}),\\
C~R_{\rm O}^{\beta},~~~~~~({\rm if}~R_{\rm O} > R_{\rm O,sat}).
\end{aligned}\right.
\end{equation} 
where $R_{\rm O,sat}$ is the Rossby number at which X-ray saturation occurs, $(L_{\rm X}/L_{\rm bol})_{\rm sat}$ is a constant indicating the saturated X-ray activity level at $R_{\rm O}\leq R_{\rm O,sat}$; $\beta$ is the index of the power law model for the unsaturated part of the X-ray activity, while $C$ is a constant. 
\citet{2022ApJ...931...45N} pointed out that there is no difference in the coronal parameters ($R_{\rm O}$ versus $\log (L_{\rm X}/L_{\rm bol})$) between single and binary stars. In Figure~\ref{fig:Ro_analysis}, we only use the $R_{\rm O}$-$\log (L_{\rm X}/L_{\rm bol})$ distribution (black open circles) for binary stars as a background.
As shown in Figure~\ref{fig:Ro_analysis}, the data of $Sample~1$ and $Sample~2$ are presented with red and blue circles, respectively. We attempted to fit these parts separately using the above model and also tried a combined fit. However, our data shows little sign of the power-law component. This leads us to focus on only the constant part represented by $(L_{\rm X}/L_{\rm bol})_{\rm sat}$. The fitting results $\log (L_{\rm X}/L_{\rm bol})_{\rm sat} = -3.14\pm0.10$ and $-3.66\pm0.10$ for $Sample~1$ and $Sample~2$ are shown in Figure~\ref{fig:Ro_analysis} as red and blue lines, respectively. An overall fit to the data of both parts yields $\log (L_{\rm X}/L_{\rm bol})_{\rm sat} = -3.40\pm0.10$. 

For EBXs overall, as the Rossby number increases from $0.01$ to $0.5$, the range of $\log (L_{\rm X}/L_{\rm bol})$ gradually widens from $[-3.5, -2.5]$ to $[-5.0, -2.5]$; there is no obvious turning point $R_{\rm O,sat}$ in the entire $R_{\rm O}$ range (see Figure~\ref{fig:Ro_analysis}). \citet{2022ApJ...931...45N} point out that for the single-star sample, $R_{\rm O,sat}$ appears around 0.19, while for the dwarf binary, it appears around 0.15. One reason that EBXs do not show $R_{\rm O,sat}$ at a similar location may be that the sample has few sources at those locations to form an effective model-fitting constraint.

\section{Discussion}\label{sec:Discu}

Based on the distribution of $\log T_{\rm eff} - \log g$, we divided the sample of EBXs into two sub-samples, $Sample~1$ ($\log g>4.0$) and $Sample~2$ ($\log g<4.0$). The former (mainly main-sequence stars) shows a positive correlation trend between temperature and $\log g$, while the latter exhibits a negative correlation trend (mainly sub-giants and giants, as well as a portion of stars about to depart from the main sequence). The $\log g-\log (L_{\rm X}/L_{\rm bol})$ and $R-\log (L_{\rm X}/L_{\rm bol})$ relationships shown in Figure~\ref{fig:feh_logg_Lx_Lbol} panel (6) ($\log g_{\rm break}=4.03\pm0.02$ within the $2 \sigma$ range of $\log g=4.0$) and Figure~\ref{fig:radius_vs_Lx_Lbol} panels (2), (4) and (6) confirm the existence of differences in the X-ray magnetic levels between these two samples, validating the necessity of this classification for studying the X-ray radiation properties of EBXs. Figures~\ref{fig:Lx_Lbol_distributions} and \ref{fig:radius_vs_Lx_Lbol} show that surface gravity maintains a clear negative correlation with both period and radius, respectively. In other words, the period, radius, and surface gravity change almost synchronously, meaning that sources with longer periods typically have larger radii and lower surface gravity.

\subsection{Relation of X-ray Emission with Period}\label{sec:period_X-ray_emisssion}

For EBXs, the component stars are generally considered to be tidally locked. Their rotational periods are the same as the systemic orbital period \citep{2008EAS....29....1M}.  Despite the changes in the stellar evolutionary process of EBXs owing to the filling of the Roche lobes, this would not result in their complete loss of single-star-like temperature and luminosity properties \citep{2005ApJ...629.1055Y}. The unbiased distribution for various parameters of close binaries collected in \citet{2019ApJS..244...43Z} suggests that the period of EBs is proportional to the radius. Our EBXs sample also suggests that an increase in the orbital period indicates an increase in the EBXs' radius and surface area. Because the bolometric luminosity is proportional to the star’s surface area, it is positively correlated with the period, as shown in Figure~\ref{fig:Lx_Lbol_distributions}, panel (2).

For the X-ray emission of EBXs, the linear correlation of the X-ray luminosity with the period has an almost equal slope to that of the bolometric luminosity, as shown in Figure~\ref{fig:Lx_Lbol_distributions}, which makes the X-ray activity level $\log (L_{\rm X}/L_{\rm bol})$ weakly negatively correlated with $\log P$. This suggests that the X-ray emission of the EBXs also originates from the stellar surface and that the X-ray luminosity is proportional to the surface area, which can be verified by the positive relationship between radius and $\log L_{\rm X}$ in Section~\ref{subsection:magnetic_R}. In this scenario, the X-ray emission of EBXs is produced by the overall surface convection zone of the star via the magnetic dynamo mechanism rather than concentrated in certain dense regions, although we cannot exclude the presence of dense active regions (e.g., spots), which could enhance the X-ray emission to some extent. Additionally, the relationships listed in Equation~\ref{equ:EBX_LX_logP} provide an empirical method to quickly determine the X-ray luminosity of an EBX based on the period, and can also be used to compare this empirical prediction with actual observed X-ray luminosity; the relationship also provides an observational constraint for the construction of a model of the magnetic activity radiation for the EBXs. 

\subsection{Relation of X-ray Emission with Surface Gravity and Radius}\label{discuss:Logg_R_X-ray_emisssion}

From the perspective of atmospheric parameters, along the direction of decreasing $\log g$, $\log g-\log L_{\rm X}$ shows an overall increasing trend. The magnetic activity levels of main-sequence components in $Sample~1$ ($\log g > 4.0$ in Figure~\ref{fig:feh_logg_Lx_Lbol}) exhibit a consistent decrease with decreasing $\log g$, while for sub-giants and giants in $Sample~2$ ($\log g < 4.0$ in Figure~\ref{fig:feh_logg_Lx_Lbol}), this trend is the opposite. From the perspective of the geometric structure of binary systems, with increasing radius, $R-\log L_{\rm X}$ also generally exhibits a positive correlation trend. For the magnetic activity level, in $Sample~1$ (triangle points in Figure~\ref{fig:radius_vs_Lx_Lbol}), the magnetic activity levels of main-sequence stars show a decrease with increasing radius, while for sub-giants and giants in $Sample~2$ (circle points in Figure~\ref{fig:radius_vs_Lx_Lbol}), this trend is the opposite. It is evident that $\log g$ and $R$ maintain a high degree of consistency with the magnetic activity properties of EBXs. These two parameters ($\log g$ and $R$) obtained independently connect atmospheric parameters with the geometric structure of EBXs, confirming that the magnetic activity of EBXs likely originates from the convection zone on the stellar surface.

\subsection{Local Structure in $\log T_{\rm eff}$-$\log L_{\rm X}$ Distribution} \label{discuss:sec:locat_structure_Teff_LX}

The distribution of color-$\log L_{\rm X}$ has been used in studies of stellar magnetic activity, such as those published by \citet{2004A&ARv..12...71G} and \citet{2022ApJ...931...45N}. In contrast, as shown in panels (4) of Figure~\ref{fig:Teff_Lx_Lbol_distributions}, we directly describe the relationship between the effective temperature $\log T_{\rm eff}$ and the X-ray luminosity $\log L_{\rm X}$.

As implied by the best-fit line in Figure~\ref{fig:Teff_Lx_Lbol_distributions} panel (4) for $Sample~1$ objects, the X-ray luminosity $\log L_{\rm X}$ generally increases with temperature until reaching $\log T_{\rm eff}\sim 3.73$ ($T_{\rm eff}\sim5400$~K). Afterward, the X-ray luminosity begins to decrease with increasing temperature.
\citet{2022ApJ...931...45N} studied the X-ray emission properties of main-sequence stars and dwarf binaries at temperatures from $\sim$3000~K to $\sim$7900~K (corresponding to $\sim$M6 to late-A type; \citealt{2000asqu.book.....C}) of the clusters Praesepe and Hyades. They suggested that single and binary stars have similar distribution characteristics in color-$\log L_{\textrm{X}}$ and color-$\log (L_{\rm X}/L_{\rm bol})$ relations. The former relationship has an overall positive correlation trend while the latter has an overall negative correlation trend. We find that the $\log T_{\rm eff}-\log L_{\rm X}$ distribution of the main-sequence stars (~4400 K to ~7900 K; ~K6 to late A type) actually has a local structure, and our observations show that the $\log L_{\textrm{X}}$ increases with temperature ($\log T_{\rm eff}\sim$3.64 to 3.73; $T_{\rm eff}\sim$4460~K to 5400~K) and then decreases ($\log T_{\rm eff}\sim$3.73 to 3.90; $T_{\rm eff}\sim$5400~K to 7950~K). 

Comparing with the color-$\log L_{\textrm{X}}$ relationship in Figure~9 of \citet{2022ApJ...931...45N}, one can identify that the X-ray luminosity of two clusters both show initial increasing trends ($\sim$M0 to $\sim$G1 type; $\sim$3840K to $\sim$5860K; \citealt{2000asqu.book.....C}) followed by decreasing trends (earlier than G1 type; $T_{\rm eff} \gtrapprox$~5860K) as the color index decreases, and the turning point of the trend occurs at the G1 spectral type with the corresponding temperature of $\sim$5860 K. Owning to the scatter of data, our results on this local trend of the X-ray luminosity of main-sequence stars with temperature are generally consistent with that of \citet{2022ApJ...931...45N}. So we find a peak-like (first up and then down) trend in the $\log T_{\rm eff}-\log L_{\rm X}$ space for the EBXs with main-sequence components for the first time.

 \subsection{Magnetic Activity versus Effective Temperature} \label{discuss:sec:Effective_Teff_X-ray_emisssion}

As shown in Figure~\ref{fig:Teff_Lx_Lbol_distributions} panels (4) and (5), for the main-sequence stars in $Sample~1$, at $\log T_{\rm eff}\sim 3.64$, the $\log (L_{\rm X}/L_{\rm bol})$ values of the EBXs distribute around the saturation level -3 of magnetic activity, and we suggest that EBXs at this temperature have the thickest surface convection zones. However, since binaries at this temperature correspond to the lowest mass and radius, the X-ray luminosity is the lowest. 
Based on the fitting of distribution $\log T_{\rm eff}- \log(L_{\rm X}/L_{\rm bol})$ (the first line in Table~\ref{table:M1_log_lx/lbol}), the magnetic activity still holds the saturation level $\leq-3.0$ until $\log T_{\rm eff}\sim 3.70$. The increase in temperature and period corresponds to objects with larger mass and radii and, hence, larger surface areas. Therefore, X-ray luminosity increases with increasing temperature.
According to the magnetic dynamo model, as the temperature increases, the convection zone becomes thinner, leading to a lower magnetic activity level. However, the larger radius and more X-ray emission area produce more X-ray radiation to compensate for the decrease of X-ray activity, which leads to an X-ray luminosity peak at the temperature $\log T_{\rm eff}\sim3.73$. As the temperature continues to increase, it reaches a point where there is insufficient material in the convection zone on its surface to maintain a typical magnetic dynamo, which is primarily powered by convection and rotation. Then, the X-ray luminosity continues to decrease because the weakening of the magnetic activity owing to the thinning of the convection zone cannot be compensated by the increase in X-ray luminosity caused by the larger stellar radius and surface area.

For the sub-giants and giants in $Sample~2$ shown in Figure~\ref{fig:Teff_Lx_Lbol_distributions} panels (6) and (7), along the direction of the arrow, when the temperature decreases, the $\log g$ decreases (color map; corresponding to the period and radius increases). Conversely, both the X-ray luminosity $\log L_{\rm X}$ and activity level $\log (L_{\rm X}/L_{\rm bol})$ increase monotonically. The above phenomenon can be explained by the decrease in temperature prompts a thickening of the convection zone. Simultaneously, the direction of temperature decrease is also the direction of increasing radius, which increases the area generating X-ray radiation. The combination of these two factors leads to an X-ray radiation trend that steadily increases as temperature decreases.

As shown in Figure~\ref{fig:Teff_Lx_Lbol_distributions} panels (3), (5), and (7), component stars of EBXs in different evolutionary stages (the main sequence, sub-giant, and giant stages) within $Sample~1$ and $Sample~2$ exhibit similar magnetic activity levels under the same temperature conditions. This may indicate that the magnetic activity in EBXs is related to temperature, and EBXs at different evolutionary stages can have similar magnetic activity levels.

In summary, based on the magnetic dynamo model, we explained the physical mechanism of the two different magnetic activity properties in sub-samples of EBX by elucidating the relationship between changes in magnetic activity level and X-ray luminosity due to variations in convection zone thickness and radiation area. The effective temperature $T_{\rm eff}$ may serve as an indicator of magnetic activity levels. For EBXs, at lower temperatures ($\log T_{\rm eff}\sim 3.64$), the magnetic activity on the stellar surface reaches saturation. In contrast, at higher temperatures, the convective layer on the stellar surface gradually becomes thinner, leading to weakened magnetic activity. We suggest that the temperature $\log T_{\rm eff}\sim 3.73$ serves as a crucial threshold for EBXs, indicating the balance between the X-ray luminosity diminishing due to the thinning of the convection zone and the increasing X-ray luminosity caused by the enlargement of the convection zone area. This also demonstrates, from the perspective of stellar structure (convection zone thickness and surface area), that EBX systems are highly significant objects for testing the stellar dynamo model.

\subsection{Magnetic Activity versus Stellar Mass} \label{discuss:sec:Stellar_mass_X-ray_emisssion}

As shown in Figures~\ref{fig:Mass1_vs_Lx_Lbol} and \ref{fig:Mass2_vs_Lx_Lbol}, along the direction of mass increasing, both the primary stars and secondary stars exhibit an initial positive correlation followed by a subsequent negative correlation with $\log L_{\rm X}$ for $Sample~1$, and a monotonic decreasing trend for $Sample~2$. Meanwhile, they all show a consistent negative correlation with $\log(L_{\rm X}/L_{\rm bol})$. Therefore, we only describe the relationships between the primary star's mass and the magnetic activity.

When the primary star's mass is at its lowest value (Figure~\ref{fig:Mass1_vs_Lx_Lbol} panel 3), EBXs in $Sample~1$ (Figure~\ref{fig:Mass1_vs_Lx_Lbol} panel 4) exhibit the highest level of magnetic activity with the thickest convection zone. However, owing to the smallest radius of the star at this point, the X-ray luminosity is at its lowest value. As the mass increases to $1.04^{+0.03}_{-0.04}~\rm M_{\rm \odot}$, the magnetic activity level continues to decrease, indicating a thinning of the convection zone and a decrease in X-ray luminosity. Nonetheless, the increased X-ray radiation due to the enlarged radius and surface area counteracted this part of the decrease, resulting in a peak in X-ray luminosity. With the $M_{1}$ further increasing, the increase in X-ray radiation from the expanded surface area is not sufficient to counterbalance the continuous weakening of magnetic activity caused by the ongoing thinning of the convection zone. This results in a sustained decrease in X-ray luminosity. For the $Sample~2$ in Figure~\ref{fig:Mass1_vs_Lx_Lbol} panel (5) and (6), as the mass of the primary star decreases, the X-ray luminosity increases. This occurs because the increased radius enlarges the radiating surface area (reflected by decreasing $\log g$ and the circles in Figure~\ref{fig:radius_vs_Lx_Lbol}), and the thickening convection zone enhances the magnetic activity level (reflected by the increasing magnetic activity level).

Additionally, since $Sample~2$ comprises components that are sub-giants and giants, as well as objects about to leave the main sequence, their radii are larger than those of main-sequence components in $Sample~1$ with the same mass and temperature, consequently resulting in higher X-ray radiation intensity. Therefore, in the $\log T_{\rm eff}-\log L_{\rm X}$ and $M-\log L_{\rm X}$ distributions, both demonstrate that the X-ray luminosity of $Sample~2$ is higher than that of $Sample~1$.

The correlations between the masses of EBXs and magnetic activity serve as a direct validation of the relationships between temperature and magnetic activity. For X-ray luminosity of $Sample~1$, the trend of initially increasing and then decreasing with mass mirrors the data distribution in $\log T_{\rm eff}-\log L_{\rm X}$. Moreover, the temperature $\sim5900^{+80}_{-120}$K of the mass break-point for the primary star corresponds closely to $T_{\rm eff, break}\sim 5400$K within a range of about $3 \sigma$, considering the scatter of data. In the case of magnetic activity level for $Sample~1$, its negative correlation trend with temperature is also replicated by stellar mass. It is evident that the observed phenomena above can be explained by a positive correlation between mass and temperature for the main-sequence components with $\log g<4.0$ in EBXs. 
For $Sample~2$, there is a mutual confirmation relationship between $\log T_{\rm eff}-\log L_{\rm X}$ and $M-\log L_{\rm X}$, as well as between $\log T_{\rm eff}-\log (L_{\rm X}/L_{\rm bol})$ and $M-\log (L_{\rm X}/L_{\rm bol})$.
More importantly, in this work, stellar mass determination is accomplished through binary spectral fitting, which is independent and unaffected by the temperature used in this study. This independently validates the physical processes discussed in Section~\ref{discuss:sec:Effective_Teff_X-ray_emisssion}.

\subsection{Comparison in $R_{\rm O}$-$\log (L_{\rm X}/L_{\rm bol})$ relationship} \label{discussion:Rossby_number}

The Rossby number $R_{\rm O}$ denotes the characteristic timescale for convective motions occurring within the star's interior. In Figure~\ref{fig:Ro_analysis}, for $R_{\rm O}$, the values in $Sample~1$ are lower than those in $Sample~2$, possibly because the sources in the former generally have shorter periods than those in the latter, indicating a relatively shorter timescale for convective motions in the star's interior. We also compare the $R_{\rm O}$-$\log (L_{\rm X}/L_{\rm bol})$ distribution of binary sample in \citet{2022ApJ...931...45N} with two sub-samples of EBXs. The saturation level of the whole binary sample in \citet{2022ApJ...931...45N} is at $\sim-2.98$, which is within $2\sigma$ compared to the magnetic activity level of $Sample~1$ ($-3.14\pm0.10$). It indicates that the distribution of the magnetic activity levels of $Sample~1$ and that of binaries in \citet{2022ApJ...931...45N} are similar. The $R_{\rm O}$-$\log (L_{\rm X}/L_{\rm bol})$ distribution of $Sample~2$ is relatively lower compared to that of $Sample~1$ by 0.52~dex, possibly because a higher fraction of high-temperature sources in $Sample~2$ leads to a relatively lower average level of magnetic activity. This makes that the combined distribution of the magnetic activity level of our full sample is lower compared to the binary sample collected by \citet{2022ApJ...931...45N}.

We suggest that the range of orbital sizes could also contribute to differences in the distribution of magnetic activity levels. The differences in the orbital sizes between the samples of \citet{2022ApJ...931...45N} and our work are evident in their respective distributions of orbital periods. The binaries in \citet{2022ApJ...931...45N} have a period distribution with 1.183, 7.955, and 13.910 days at the 16th, 50th, and 84th percentiles, respectively, while our sources have the period value of 0.403, 0.597, and 1.364 days at the corresponding percentiles. The former sample has substantially more binaries with wider orbits. \citet{2022ApJ...931...45N} indeed suggested that the components of their binaries may lack interaction, while there is no doubt that complex material exchanges and transfers occur in EBXs \citep{2005ApJ...629.1055Y, 2011ASPC..451...25Z}. Therefore, we infer that the different degrees of matter transfer or exchange may affect the magnetic activity. Further confirmation of this inference will require the collection of additional samples and detailed analyses of individual systems in the future.

\subsection{Comparison with EWXs} \label{subsec:EBX_and_EWX}

The W Ursa Majoris (W UMa-type) binary, also referred to as an EW-type binary, is characterized by a contact configuration in which both components fill their Roche lobes and jointly share a common envelope. The EW-type binaries with X-ray emission (EWXs; e.g., \citealt{2001A&A...370..157S, 2004A&A...415.1113G, 2006AJ....131..990C, 2019PASP..131h4202L, 2022A&A...663A.115L}) also constitute an important type of X-ray sources. Additionally, EB-type binaries are generally considered to be precursor stars to EW-type binaries \citep{2005ApJ...629.1055Y}. Therefore, it is worth comparing the X-ray radiation properties of EBXs and EWXs.

The orbital period of EBXs in our sample ranges from $\sim$0.2 to $\sim$10~days. The X-ray and bolometric luminosities cover ranges of are $\sim$1.74$\times 10^{29}$ to $\sim$1.32$\times 10^{32}$erg~s$^{-1}$, and $\sim$5.20$\times 10^{32}$erg s$^{-1}$ to $\sim$3.98$\times 10^{36}$erg~s$^{-1}$, respectively. The lower limits of luminosity are close to those of the EWXs studied by \citet{2022A&A...663A.115L}, while the upper limits are approximately two orders of magnitude higher than those of the EWXs. However, if we limit the EBXs in the same period ranges as the EWXs (0.2 to 0.44 days), these physical parameters of EBXs are almost similar to those of EWXs \citep{2022A&A...663A.115L}.

EBXs and EWXs have similar qualitative correlations between the stellar spectral parameters (temperature, metallicity, and surface gravity) and X-ray emissions (luminosity and activity level), except for the $\log T_{\rm eff}$-$\log L_{\rm X}$ relation (see Section~\ref{subsubsec:Teff} in this work and Section 3.2 of \citealt{2022A&A...663A.115L}). For this particular relationship, the positive correlation trend of both EBXs and EWXs in the temperature range of 4500~K to 6300~K \citep{2022A&A...663A.115L} and the range of X-ray luminosity $10^{29.5}-10^{30.5}$erg s$^{-1}$ is almost identical. Furthermore, if more EWXs with effective temperatures greater than 6300~K  are found, their X-ray luminosity will likely show a downward trend similar to what EBXs have exhibited by EBXs. Both EBXs and EWXs show a negative $\log T_{\rm eff}$-$\log (L_{\rm X}/L_{\rm bol})$ correlation, that is, objects with low temperatures have a higher magnetic activity level.

The statistical distributions of effective temperatures of both EBXs and EWXs peak at near 5600 K, while the former does not have a significant cutoff at the high temperature ($\sim$6300K) end. The metallicity distribution of EBXs peaks at -0.25~dex, which is lower compared to that of the EWXs ($\sim-0.05$~dex). For the distributions of surface gravity, both EBXs and EWXs have similar peak locations, while the former has a long tail in the interval from 3.0~dex to 3.75~dex. The EBXs present a better laboratory for testing the magnetic dynamo model because it reflects the magnetic activity characteristics of main-sequence stars and sub-giant and giant stars, and the similarities and differences between them as discussed in the above sections, while the EWXs only mainly include the main-sequence stars as components that have strong positive correlations between period, mass, and temperature \text{at $P\lesssim0.44~days$}.

\section{Summary}\label{sec:Summary}

Based on the AVSD database, we collected the X-ray counterparts of 255 EB-type binaries from the \textit{XMM-Newton}, RASS and \textit{Chandra} databases. Correlation analyses of the period and spectral parameters (i.e., effective temperature, metallicity, surface gravity, the masses and radii of component stars) with the X-ray emission properties were performed for the first time for EBXs. Based on the $\log g-\log T_{\rm eff}$ distribution, we divided the EBXs into $Sample~1$ with $\log g>4.0$ and $Sample~2$ with $\log g<4.0$. The former, which is primarily composed of main-sequence member stars, as the temperature increases, $\log g$ decreases, and the stellar radius increases. On the other hand, the latter, mainly composed of sub-giants, giants, and a portion of stars about to depart from the main sequence, exhibits the characteristic that as the temperature decreases, $\log g$ decreases, and the stellar radius increases. The main conclusions are as follows:

\begin{enumerate}

\item The X-ray and bolometric luminosity both increase with longer orbital periods. The rates of change are consistent, indicating that the increases in the X-ray and bolometric luminosity of the EBXs are almost synchronous along the period. The X-ray emission may originate from the convection zones of EBXs. Both parameters ($\log L_{\rm X}$ and $\log L_{\rm bol}$) are positively correlated with the surface area of the binary system.

\item Among the atmospheric parameter, surface gravity $\log g$ and X-ray luminosity show a strong negative correlation, while the distribution of $\log g - \log(L_{\rm X}/L_{\rm bol})$ can be described using a segmented linear fit, where the level of magnetic activity of the EBXs is proportional to the surface gravity at $\log g>4.0$, and the opposite at $\log g<4.0$. $\log g$ (in decreasing direction) exhibits consistency with the binary geometric parameter $R$ in the relationships with $\log L_{\rm X}$ and $\log(L_{\rm X}/L_{\rm bol})$. The results of both validate that the X-ray radiation of EBXs likely originates from the convection zone. In addition, metallicity [Fe/H] is almost independent of the X-ray emission.

\item  We found that the X-ray emission luminosity of the main-sequence components with $\log g>4.0$ in EBXs shows an increasing and then decreasing trend with the effective temperature, and confirmed these distributions by comparison with binaries in the Paesepe and Hyadea clusters.

\item We found for the first time the differences in magnetic activity properties for EBXs in different evolution stages (main-sequence, sub-giant, and giant stages). These differences are reflected in the $\log T_{\rm eff}$-$\log L_{\textrm{X}}$, $\log T_{\rm eff}$-$\log(L_{\textrm{X}}$/$L_{\textrm{bol}})$, $\log g$-$\log(L_{\textrm{X}}$/$L_{\textrm{bol}})$, $M$-$\log L_{\textrm{X}}$, $M$-$\log(L_{\textrm{X}}$/$L_{\textrm{bol}})$ and $R$-$\log(L_{\textrm{X}}$/$L_{\textrm{bol}})$ relationships.
Based on the magnetic dynamo model, we used changes in the surface convection zone area and temperature-induced changes in convection zone thickness to explain the physical origin. We suggest $\log T_{\rm eff}\sim 3.73$ as a crucial temperature value for EBX in testing the magnetic dynamo model because, at this temperature, a balance in X-ray luminosity is achieved due to the combined influence of variations in the thickness and surface area of the convection zone. We found a strong negative correlation between the temperature and the magnetic activity level of $\log(L_{\rm X}/L_{\rm bol})$. A higher temperature leads to a thinner convection zone, and thus weaker magnetic activity. Furthermore, the magnetic activity levels may be related to the temperature and mass of EBX.

%%%%%######independent of its evolutionary stage.

\item We developed the single- and binary-star spectral model to fit the spectra of LAMOST DR9 and then derive the masses and radii for the primary and secondary components in EBXs. The mass versus magnetic activity ($\log L_{\rm X}$ and $\log(L_{\rm X}/L_{\rm bol})$) exhibits distributions similar to those observed in the effective temperature versus magnetic activity. Both mutually support the explanation of the relationship between the thickness of the convection zone and the surface area in EBX's X-ray activity. While mass provides insights into the physical essence, samples with effective temperature are more numerous and statistically significant.

\item Regarding the $R_{\rm O}$-$\log(L_{\textrm{X}}/L_{\textrm{bol}})$ relation of EBXs, it is difficult to constrain with a constant plus power-law model. The $R_{\rm O}$-$\log(L_{\textrm{X}}/L_{\textrm{bol}})$ distribution for EBXs in $Sample~1$ is consistent with the binaries in Praesepe and Hyade clusters within a 1$\sigma$ range. The $Sample~1$ has a shorter timescale for convective motions compared to $Sample~2$ and has a higher magnetic activity level, which might be due to the former having more sources with short periods and low temperatures. The overall distribution of EBXs in this relationship is lower than that of binaries in clusters Praesepe and Hyades, which may result from the fact that the EBX sample contains sources with different material exchange or transfer rates.

\item The X-ray luminosity and activity levels of the EBXs were consistent with those of the EWXs at similar periods and temperatures. Because EBXs cover a wider range of periods and spectral parameters, they provide an important laboratory for studying magnetic generator mechanisms.

\end{enumerate}

\section*{Acknowledgements}

We are grateful to the anonymous referee for providing thoughtful and helpful comments that improved the manuscript.
We appreciate the discussions on stellar evolution with Chun-Yan Li from SHAO, CAS.
This work is supported by the National Natural Science Foundation of China (NSFC) under grant numbers 12273029, U1938105, and 12221003. 
We acknowledge the data support from Guoshoujing Telescope. Guoshoujing Telescope (the Large Sky Area Multi-Object Fiber Spectroscopic Telescope; LAMOST) is a National Major Scientific Project built by the Chinese Academy of Sciences. Funding for the project has been provided by the National Development and Reform Commission. LAMOST is operated and managed by the National Astronomical Observatories, Chinese Academy of Sciences.
We also acknowledge the support of X-ray data based on observations obtained with \textit{XMM-Newton}, an ESA science mission with instruments and contributions directly funded by ESA Member States and NASA. This work has made use of data from the European Space Agency (ESA) mission
{\it Gaia} (\url{https://www.cosmos.esa.int/gaia}), processed by the {\it Gaia}
Data Processing and Analysis Consortium (DPAC,
\url{https://www.cosmos.esa.int/web/gaia/dpac/consortium}). Funding for the DPAC
has been provided by national institutions, in particular, the institutions
participating in the {\it Gaia} Multilateral Agreement. This paper employs a list of Chandra datasets, obtained by the Chandra X-ray Observatory, contained in~\dataset[DOI:10.25574/cdc.195]{https://doi.org/10.25574/cdc.195}.

\appendix\label{appendix}

\section{Different Marginalized posterior probability distributions}
\label{app:posterior_distributions}

\begin{figure}[!htb]
\centering
\includegraphics[width = 9cm]{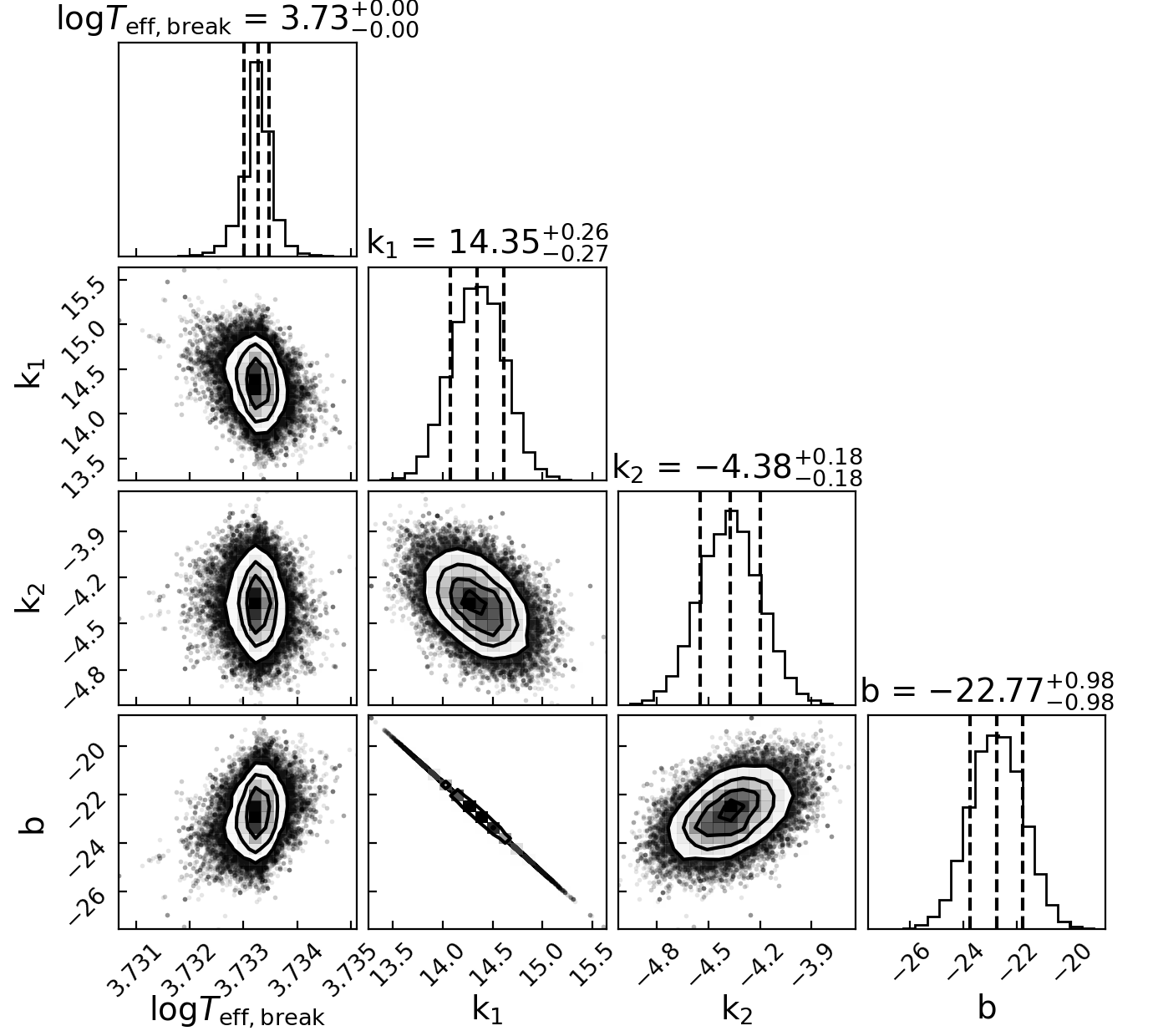}
\caption{Marginalized posterior probability distributions from the MCMC fitting for the $\log T_{\rm eff}$-$\log L_{\textrm{X}}$ relationship of $Sample~1$. The parameter values are the peaks of the one-dimensional distributions, while the vertical dashed lines are located at the 16th, 50th, and 84th percentiles.}
\label{fig:Teff_Lx_Lbol_corner}
\end{figure}

\begin{figure}[!htb]
\centering
\includegraphics[width = 9cm]{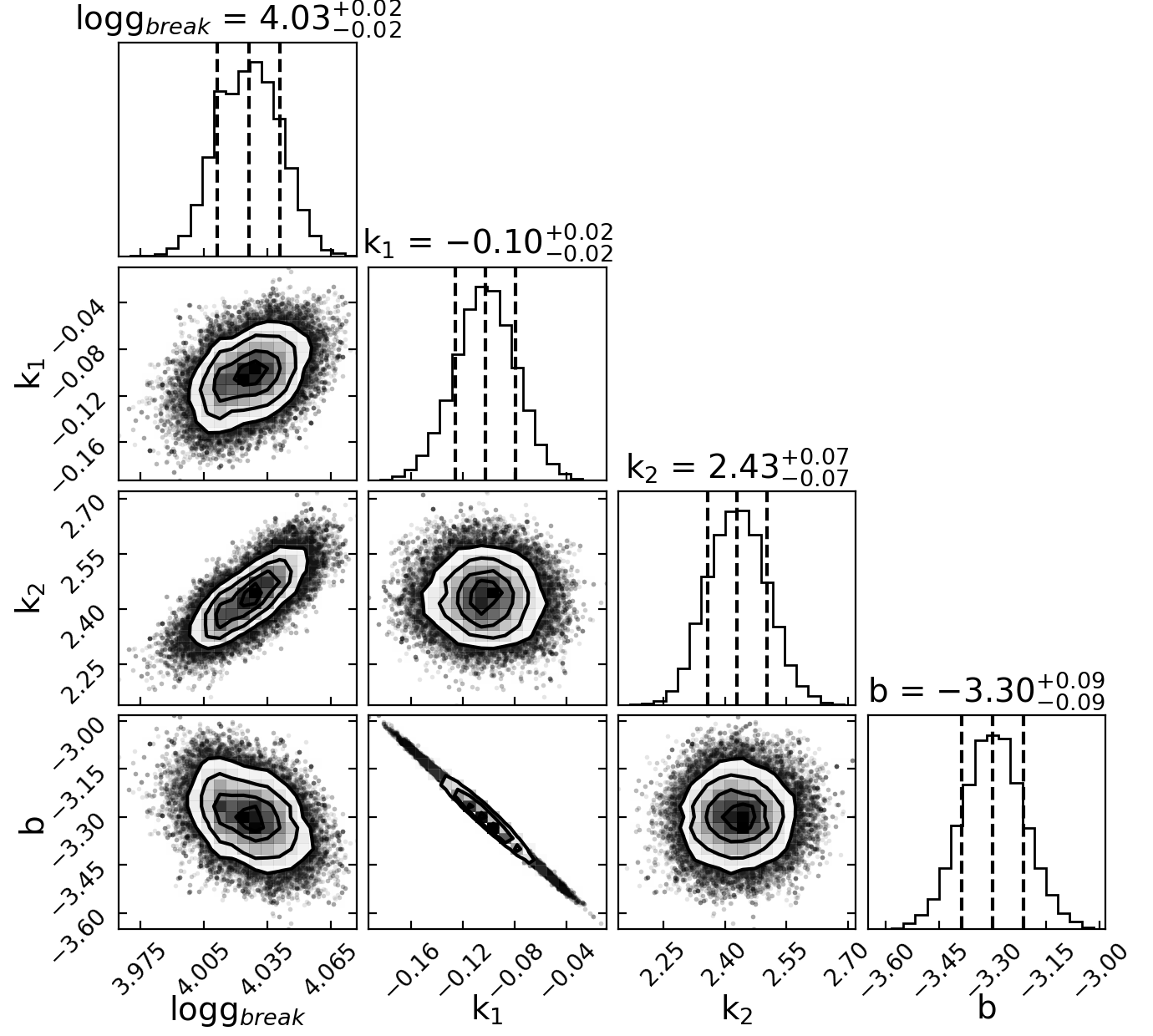}
\caption{Marginalized posterior probability distributions from the MCMC fitting for the $\log g$-$\log (L_{\textrm{X}}/L_{\textrm{bol}})$ relationship. The parameter values are the peaks of the one-dimensional distributions, while the vertical dashed lines are located at the 16th, 50th, and 84th percentiles.}
\stepcounter{appendixfigure}
\label{fig:feh_logg_Lx_Lbol_corner}
\end{figure}

\begin{figure}[!htb]
\centering
\includegraphics[width = 8cm]{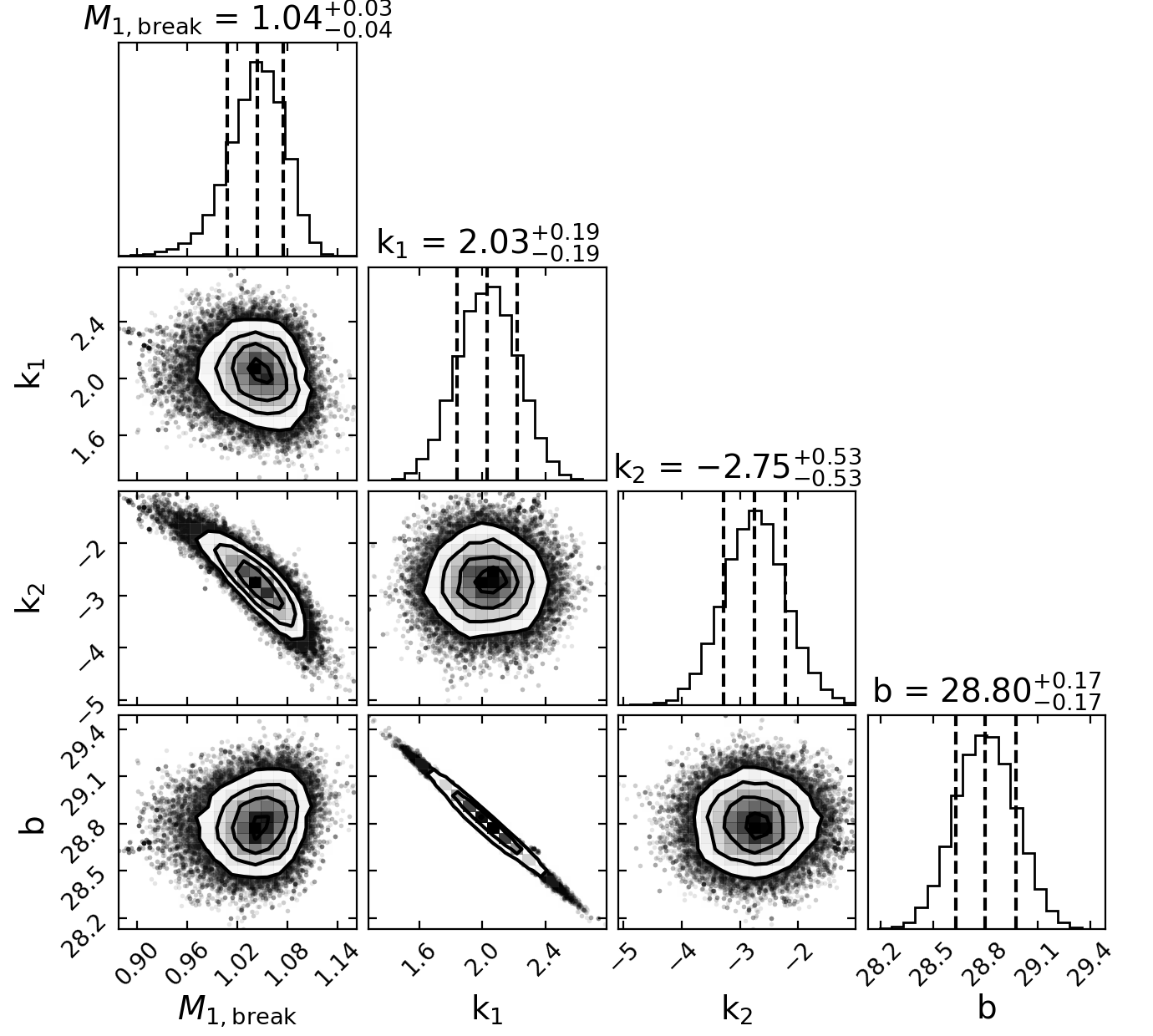}
\includegraphics[width = 8cm]{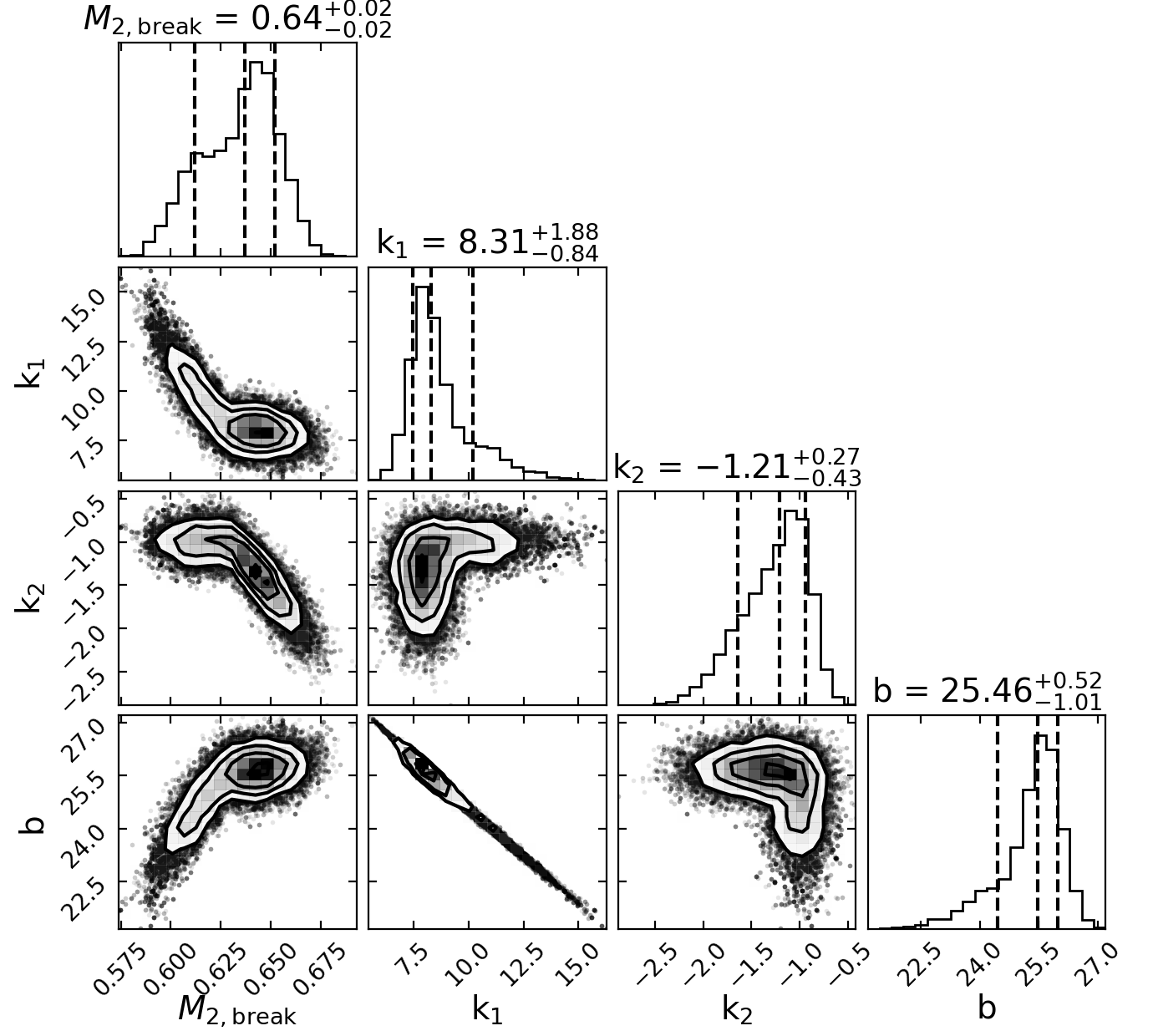}
\caption{Marginalized posterior probability distributions from the MCMC fitting for the $M_{1}$-$\log L_{\textrm{X}}$ (left) and $M_{2}$-$\log L_{\textrm{X}}$ (right) relationships. The parameter values are the peaks of the one-dimensional distributions, while the vertical dashed lines are located at the 16th, 50th, and 84th percentiles.}
\stepcounter{appendixfigure}
\label{fig:M1_Lx_Lbol_corner}
\end{figure}

\begin{figure}[!htb]
\centering
\includegraphics[width = 8cm]{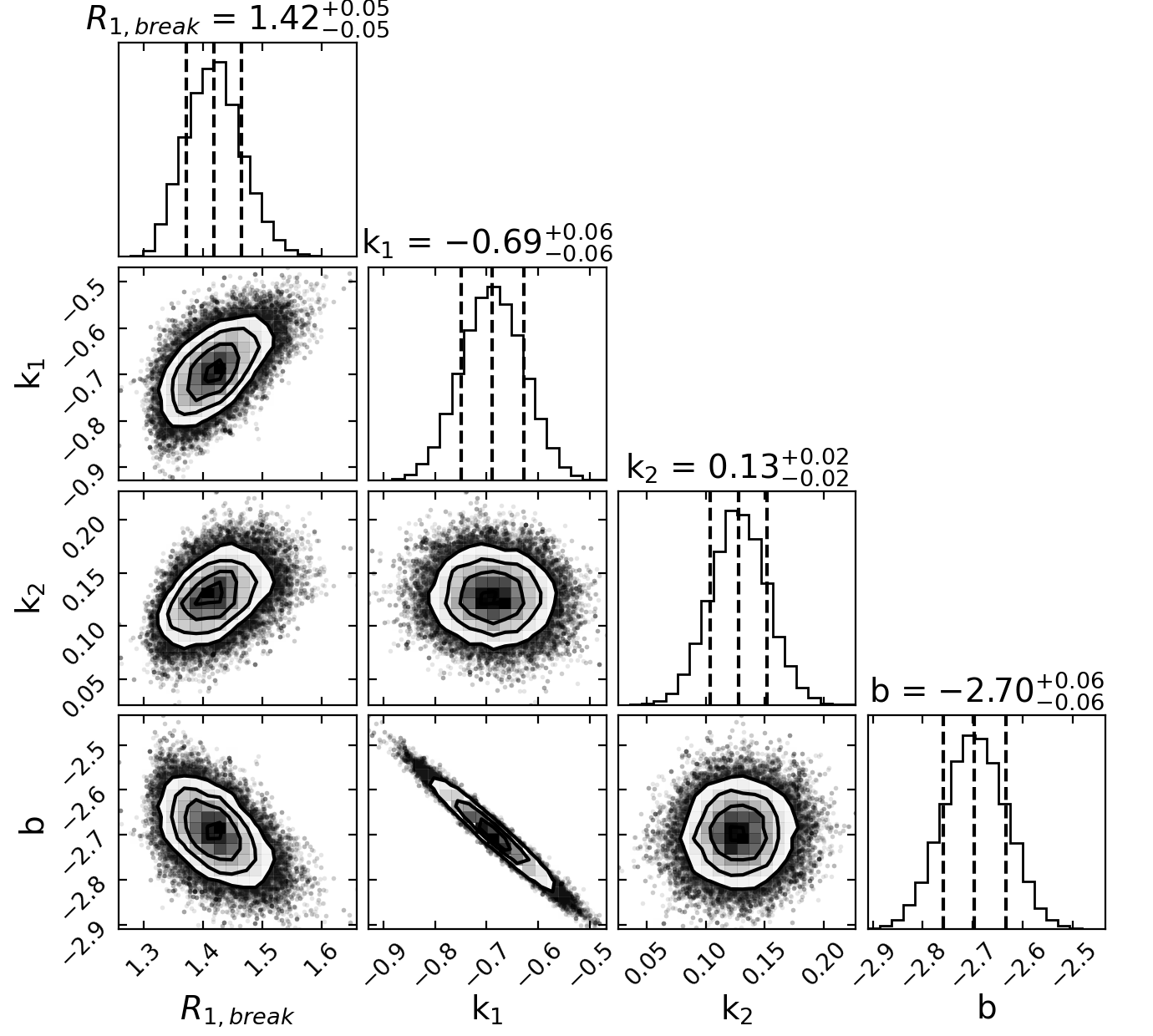}
\includegraphics[width = 8cm]{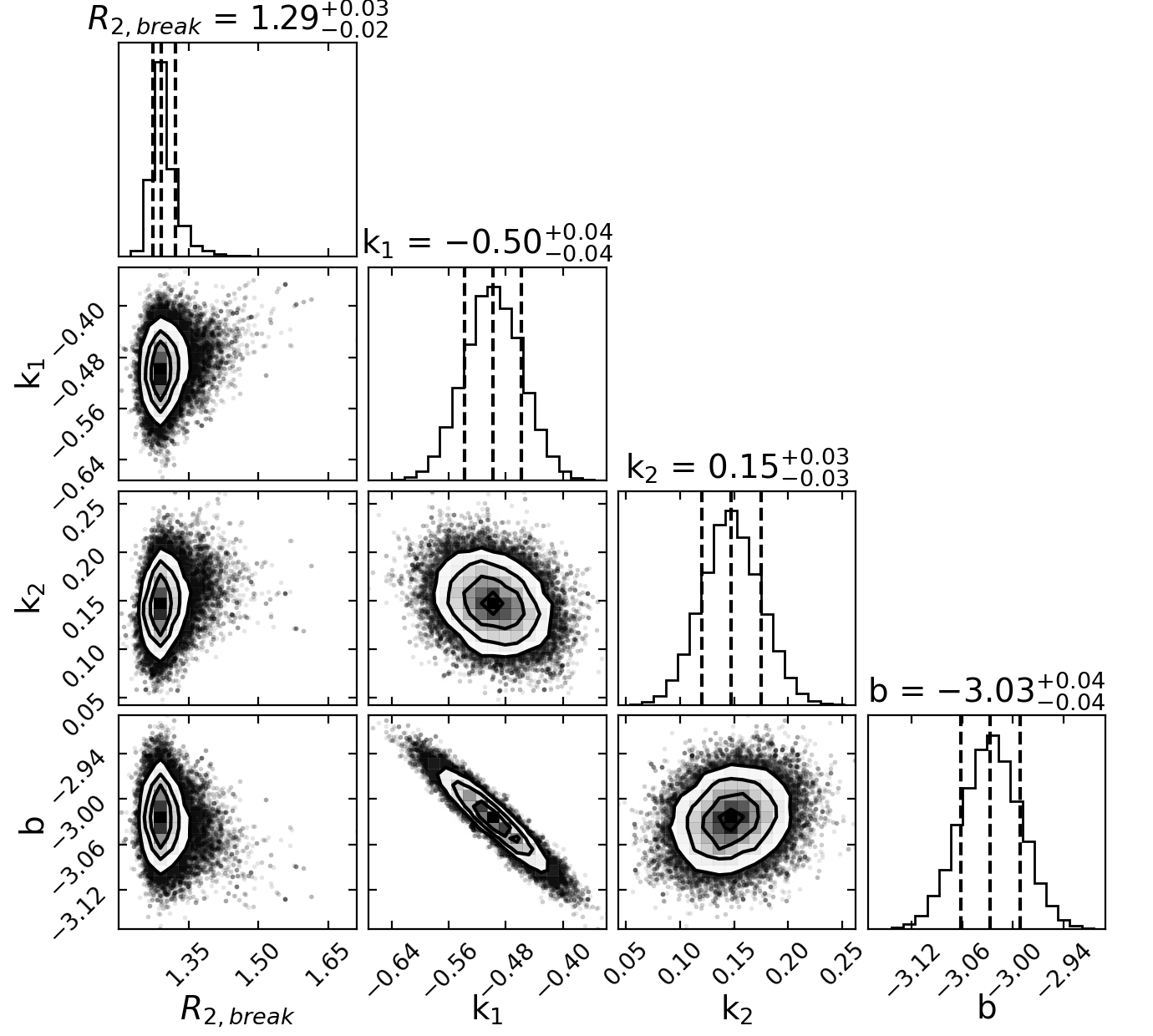}
\includegraphics[width = 8cm]{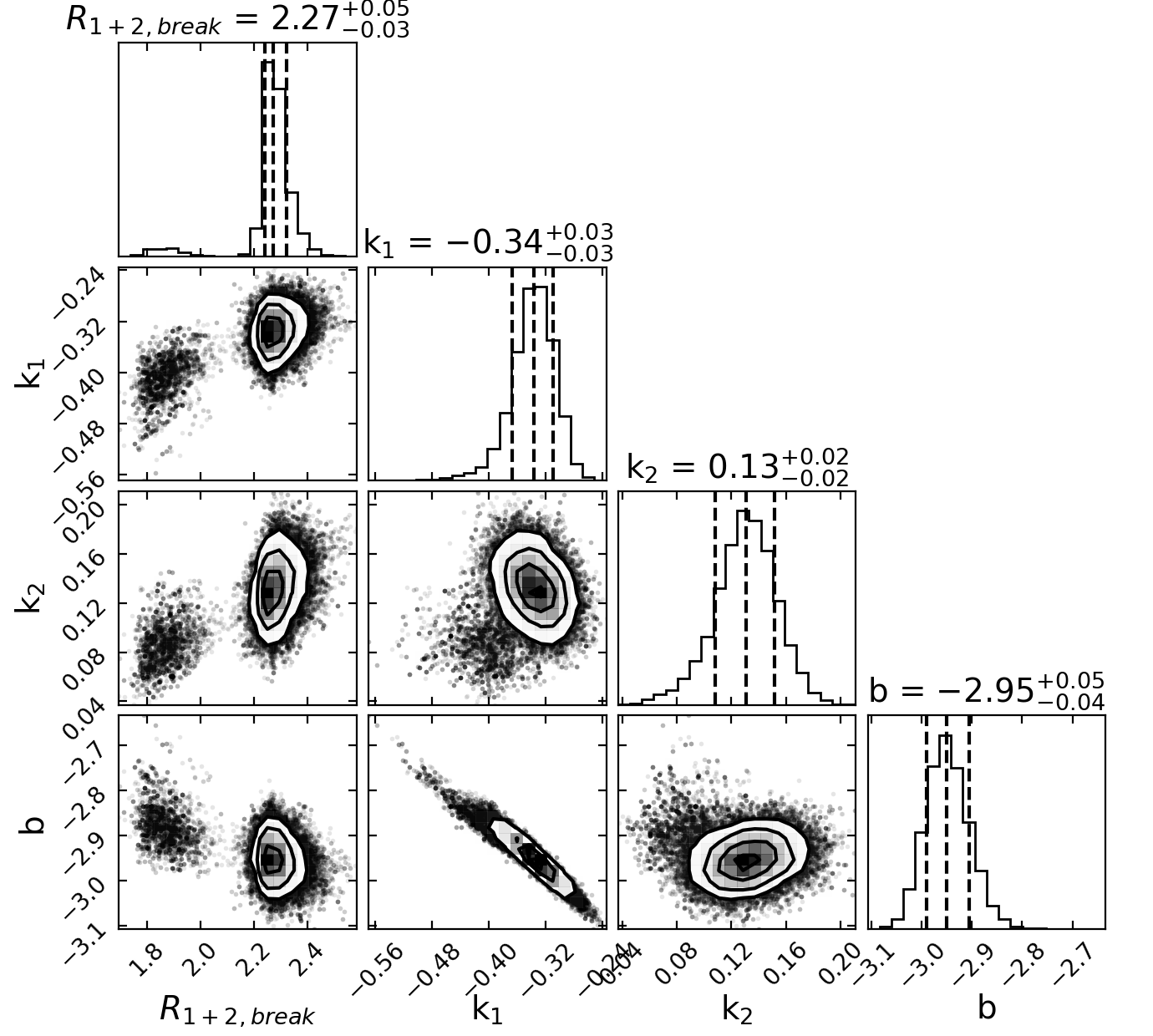}
\caption{Marginalized posterior probability distributions from the MCMC fitting for the $R_{1}$-$\log (L_{\textrm{X}}/L_{\textrm{bol}})$ (upper left), $R_{2}$-$\log (L_{\textrm{X}}/L_{\textrm{bol}})$ (upper right), and  $R_{1+2}$-$\log (L_{\textrm{X}}/L_{\textrm{bol}})$ (lower) relationships. The parameter values are the peaks of the one-dimensional distributions, while the vertical dashed lines are located at the 16th, 50th, and 84th percentiles.}
\stepcounter{appendixfigure}
\label{fig:R1_Lx_Lbol_corner}
\end{figure}

\bibliography{reference}{}
\bibliographystyle{aasjournal}

\end{document}